\documentclass[12pt]{article}

\usepackage{graphicx}
\usepackage{amsmath}
\usepackage{amssymb}

\numberwithin{equation}{section}

\usepackage{color}
\input{colordvi.tex}

\newcommand{\ket}[1]{\left|{#1}\right>}
\newcommand{\vev}[1]{\left<{#1}\right>}
\newcommand{\vac}[1]{\left|{0}\right>}

\begin{document}

\begin{titlepage}
\thispagestyle{empty}
\begin{flushleft}
RIKEN-TH-125
\hfill April, 2008 \\
UT-08-07
\end{flushleft}

\vskip 1.5 cm
\bigskip

\begin{center}
{\Large \bf{
Gauge Invariant Overlaps for Classical Solutions 
}}\\
\vskip .2in
{\Large\bf{ in Open String Field Theory}}\\

\renewcommand{\thefootnote}{\fnsymbol{footnote}}

\vskip 2.5cm
{\large
Teruhiko Kawano$^{a}$,
Isao Kishimoto$^{b}$ 
and Tomohiko Takahashi$^{c}$}\\

\bigskip\bigskip

{\it
$^{a)}$ Department of Physics, University of Tokyo,
Hongo, Tokyo 113-0033, Japan\\
\smallskip
$^{b)}$ Theoretical Physics Laboratory, RIKEN,
Wako 351-0198, Japan\\
\smallskip
$^{c)}$ Department of Physics, Nara Women's University,
Nara 630-8506, Japan
}
\vskip 2in
\end{center}
\begin{abstract}

We calculate gauge invariant observables for 
Schnabl's solution with analytic method and 
with the level truncation approximation. 
We also compute them for the numerical solution 
initially obtained by Sen and Zwiebach 
in the level truncation approximation to 
compare with the one for Schnabl's solution. 
The results are consistent with the expectation 
that they may be gauge equivalent. 
We briefly discuss the gauge invariant observables and the action 
for a marginal solution with a nonsingular current.

\end{abstract}
\end{titlepage}\vfill\setcounter{footnote}{0} \renewcommand{\thefootnote}{
\arabic{footnote}} \newpage

\section{Introduction}

After the advent of the paper \cite{Sen:1999nx}, tachyon condensation 
has been studied in Witten's bosonic open string field theory. 
They used the level truncation approximation \cite{K-S} to obtain 
a classical solution numerically  and obtained its vacuum energy giving 
99\% of the D-brane tension. The approximation have been improved 
soon later to the higher level \cite{Moeller:2000xv,Gaiotto:2002wy}. 

More recently, Schnabl has discovered an analytic classical solution 
in the open string field theory. The analytic solution also describes 
tachyon condensation, because its vacuum energy is exactly equal to
the D25-brane tension \cite{Schnabl_tach,Okawa:2006vm,Fuchs:2006hw}, 
and the BRST cohomology around it is trivial \cite{ES}, 
which implies that there are no physical degrees of freedom 
of open strings perturbatively, as conjectured by Sen 
\cite{Sen:1999mh,Sen:1999mg,Sen:1999xm}.

We thus have two classical solutions of the open string field theory; 
the numerical one and the analytical one. Therefore, it seems natural 
to ask the relation between them. Since they give the same vacuum 
energy equal to the D-brane tension, one likely suspects that they may be 
gauge equivalent solutions. In order to confirm the expectation, one needs 
gauge invariant observables to compare their values for these two solutions. 

Zwiebach introduced the couplings of a single on-shell closed string state
with open string field into the open string field theory 
in a gauge invariant fashion \cite{Zwiebach:1992bw}. 
In fact, it has been shown \cite{Takahashi:2003kq,Garousi:2004pi} that 
the couplings exactly give the disk amplitudes with two closed strings 
and the ones with one closed string and two open strings.
They have also been used to discuss the closed string degrees of freedom 
in vacuum string field theory \cite{Gaiotto:2001ji}. 
The gauge invariance of the coupling with the closed string tachyon 
was reassured \cite{Hashimoto:2001sm} in terms of the oscillator 
formalism. Hashimoto and Itzhaki also emphasized that they are also 
gauge invariant observables \cite{Hashimoto:2001sm}. 

In this paper, the gauge invariant observables will be called 
gauge invariant overlaps to distinguish with general 
gauge invariant observables. 

In this paper, we will calculate the gauge invariant overlaps for 
Schnabl's solution $\Psi_{\lambda}$ with the parameter $\lambda$ 
analytically in the sliver frame and numerically with the level 
truncation. The results on the vacuum energy 
of the solution $\Psi_{\lambda}$ imply that 
only the solution with $\lambda=1$ cannot be gauged away and 
is thus physically non-trivial, while the rest is all trivial. 
The analytical results on the gauge invariant overlaps for 
the solution $\Psi_{\lambda}$ are consistent with the ones 
of the vacuum energy. There however exist subtleties in the evaluation, 
and thus we will confirm the results numerically in the level 
truncation approximation, as done for the vacuum energy in 
\cite{Schnabl_tach,Takahashi:2007du}

We will also compute the gauge invariant overlaps for the numerical 
solution $\Psi_{\rm N}$ in \cite{Sen:1999nx,Moeller:2000xv,Gaiotto:2002wy} 
with the level truncation to compare with the results for 
Schnabl's solution with $\lambda=1$, and we will show that the comparison 
is consistent with the expectation that they are gauge equivalent.

Furthermore, we will briefly report our results on the gauge invariant 
overlaps for a marginal solution with a nonsingular current $i\partial X^+$ 
in \cite{Schnabl:2007az, Kiermaier:2007ba}
and  that it vanishes as an evaluation of the action. 

This paper is organized as follows. 
In the next section, we will give a brief review on the gauge invariant 
overlaps of the on-shell closed string states
and make a few comments on the relation to the open-closed string vertex. 
In \S \ref{sec:GinvSch}, we will evaluate 
the gauge invariant overlaps for Schnabl's solution $\Psi_{\lambda}$, 
both analytically and numerically.
In \S \ref{sec:Siegel}, we will give the results for the numerical solution 
$\Psi_{\rm N}$. In \S \ref{sec:GinvMarg}, we will discuss the gauge invariant 
for the marginal solution analytically.
In \S \ref{sec:Dis}, we give summary and discussions on our results.
In appendices, we will explicitly give a few examples of the gauge invariant 
overlaps and the Shapiro-Thorn's open-closed string vertex, 
and will explain technical details on our computations.

\section{
Gauge Invariant Overlaps 
\label{sec:On_Gauge}}

In Witten's bosonic open string field theory, the action is given by 
\begin{eqnarray}
S[\Psi]&=&-{1\over g^2}\left({1\over 2}\langle \Psi,
Q_{\rm B} \Psi\rangle+{1\over 3}\langle \Psi,\Psi*\Psi\rangle\right),
\end{eqnarray}
which is left invariant under the gauge transformation  
\begin{equation}
\delta_{\Lambda}\Psi=Q_{\rm B}\Lambda+\Psi*\Lambda-\Lambda*\Psi
\label{gaugetrf}
\end{equation}
with a gauge `parameter' $\Lambda$.
It was discussed in \cite{Zwiebach:1992bw, Hashimoto:2001sm,Gaiotto:2001ji} 
that 
\begin{eqnarray}
 {\cal O}_V(\Psi)&=&\langle V(i)f_{\cal I}[\Psi]\rangle\,,~~~~~
f_{\cal I}(z)\equiv \frac{2z}{1-z^2}\,.
\label{eq:Zwiebach_inv}
\end{eqnarray}
is gauge invariant under (\ref{gaugetrf}), 
where the CFT correlator is defined on the upper half plane, and 
will be called a gauge invariant overlap in this paper. 
The operator $V(i)$ is inserted at the midpoint of the string field $\Psi$ 
and is a primary field of conformal dimension $(0,0)$ 
and the ghost number two. 
The holomorphic function $f_{\cal I}(z)$ maps the unit half disk 
to the upper half plane.
Since the identity state $\langle {\cal I}|$ may be defined by 
$\langle {\cal I}|\phi\rangle=\langle f_{\cal I}[\phi(0)]\rangle$, 
one may rewrite the gauge invariant overlap ${\cal O}_V(\Psi)$ as
\begin{eqnarray}
  {\cal O}_V(\Psi)&=&\langle {\cal I}|V(i)|\Psi\rangle=\langle
   \Phi_V,\Psi\rangle\,,
\label{eq:O_VPsi}
\end{eqnarray}
with the definition 
\begin{eqnarray}
{}|\Phi_V\rangle&\equiv&V(i)|{\cal I}\rangle.
\label{eq:Phi_VIket}
\end{eqnarray}
A few examples of $|\Phi_V\rangle$ in terms of the oscillators 
are explicitly given in appendix \ref{sec:Ginv_closed}.
Because $V(i)$ is inserted at the midpoint of the identity state 
$|{\cal I}\rangle$, one can see that 
$\langle \Phi_V,\Psi*\Lambda\rangle=
\langle \Phi_V,\Lambda*\Psi\rangle$, or in other words, 
\begin{eqnarray}
  {\cal O}_V(\Psi*\Lambda)={\cal O}_V(\Lambda*\Psi), 
\label{eq:com_on_closed}
\end{eqnarray}
at least na\"ively. Therefore,
by requiring $\Phi_V$ to satisfy that $ Q_{\rm{B}}|\Phi_V\rangle=0$, 
one can finds that ${\cal O}_V(\delta_{\Lambda}\Psi)=0$.
One thus obtains the gauge invariant observables ${\cal O}_V(\Psi)$. 
In particular, note that it vanishes for
pure gauge solutions 
\begin{eqnarray}
&&{\cal O}_V(e^{-\Lambda}Q_{\rm B}e^{\Lambda})=\int_0^1 dt\, {\cal
 O}_V(e^{-t\Lambda}*Q_{\rm B}\Lambda *e^{t\Lambda})
={\cal O}_V(Q_{\rm B}\Lambda)=-\langle Q_{\rm B}\Phi_V,\Lambda\rangle=0\,.
\end{eqnarray}

\begin{figure}[h]
\centerline{\includegraphics[width=11cm]{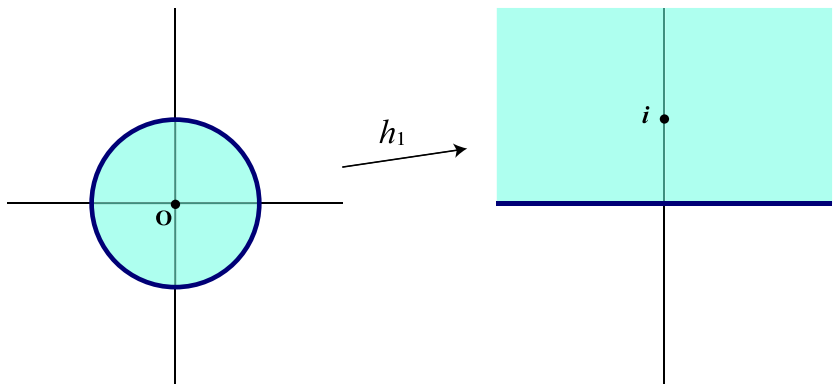}}
\caption{The conformal map $h_1$ in the definition of the Shapiro-Thorn 
vertex. The map $h_1$ transforms a unit disc to the upper half plane. The
center at the origin, the insertion point of a closed string is mapped
to $i$.} 
\label{fig:h1}
\end{figure}

\begin{figure}[h]
\centerline{\includegraphics[width=11cm]{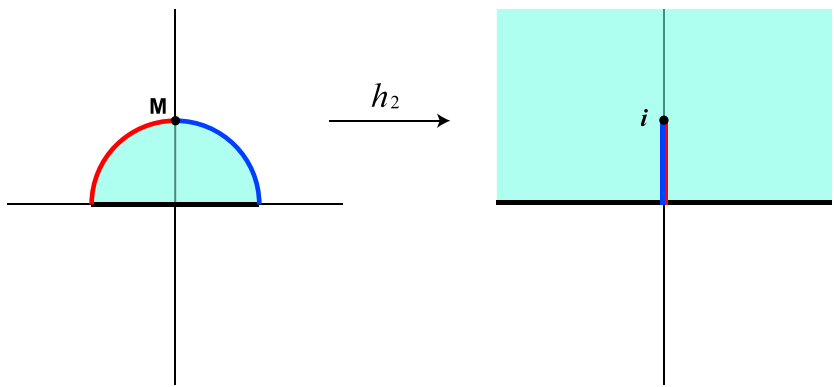}}
\caption{The conformal map $h_2$ in the definition of the Shapiro-Thorn 
vertex. The map $h_2$ transforms a unit half disc to the upper half
plane. The open string midpoint $i$ is also mapped to $i$. The left half 
and the right half of the open string are identified on the mapped plane.}
\label{fig:h2}
\end{figure}

Incidentally, it could be helpful to rewrite the gauge invariant overlap 
${\cal O}_V(\Psi)$ by using open-closed string vertex,
which maps states in the closed string Hilbert space to the ones in 
the open string one. Let us consider the Shapiro-Thorn vertex  
$\langle\hat\gamma(1_{\rm c},2)|$ given in \cite{Shapiro:1987ac},
which is defined with the conformal maps $h_1(w)=-i(w-1)/(w+1)$ 
and $h_2(w)=(w-1/w)/2$ by  
the CFT correlator on the upper half plane as 
\begin{eqnarray}
\langle\hat\gamma(1_{\rm c},2)|\phi_{\rm c}\rangle_{1_{\rm c}}|\psi\rangle_2=
\langle h_1[\phi_{\rm c}(0,0)]h_2[\psi(0)]\rangle,
\label{eq:S-T_LPPdef}
\end{eqnarray}
where 
$|\phi_{\rm c}\rangle_{1_{\rm c}}=\phi_{\rm c}(0,0)|0\rangle_{1_{\rm c}}$ 
is in the closed string Hilbert space
and $|\psi\rangle_2=\psi(0)|0\rangle_2$ is in the open string one.
These conformal maps are depicted in Figure~\ref{fig:h1} and \ref{fig:h2}. 
(See, appendix \ref{sec:ST-vertex} for detail.) 

Since $h_1(0)=i$ and $h_2(w)=I\circ f_{\cal I}(w)$, where $I(z)=-1/z$ is the
inversion map, one obtains 
\begin{eqnarray}
\langle \hat{\gamma}(1_{\rm c},2)|V_{\rm c}\rangle_{1_{\rm c}}|\psi\rangle_2
&=&\langle h_1[V_{\rm c}(0,0)]h_2[\psi(0)]\rangle
=\langle I\circ h_1[V_{\rm c}(0,0)]\, f_{\cal I}[\psi(0)]\rangle
\nonumber\\
&=&\langle V_{\rm c}(i,-i)\, f_{\cal I}[\psi(0)]\rangle~~~~~~
\end{eqnarray}
for a primary field $V_{\rm c}(z,\bar z)$ of conformal dimension $(0,0)$ and
the ghost number two.
By identifying $V(i)$ in (\ref{eq:Zwiebach_inv}) with $V_{\rm c}(i,-i)$, 
one finds that 
\begin{eqnarray}
{\cal O}_V(\Psi)&=& \langle
 \hat{\gamma}(1_{\rm c},2)|V_{\rm c}\rangle_{1_{\rm c}}|\Psi\rangle_2\,.
\label{eq:O_Vgamma}
\end{eqnarray}
It means that $|\Phi_V\rangle$ in 
(\ref{eq:Phi_VIket}) can be given by  $|V_{\rm c}\rangle=V_{\rm
{c}}(0,0)|0\rangle$:
\begin{eqnarray}
{}|\Phi_V\rangle_3 =V(i)|{\cal I}\rangle_3
=\langle\hat{\gamma}(1_{\rm c},2)|V_{\rm
  c}\rangle_{1_{\rm c}}|R(2,3)\rangle\,,
\label{eq:Phi_V3}
\end{eqnarray}
where $|R(2,3)\rangle$ is the reflector and can be used to define the BPZ
conjugation. Using the BRST invariance of the string vertices 
$\langle\hat{\gamma}(1_{\rm c},2)|$ and $|R(2,3)\rangle$
\begin{eqnarray}
 &&\langle\hat{\gamma}(1_{\rm c},2)|(Q_{\rm B}^{(1)}+\bar Q_{\rm
  B}^{(1)}+Q_{\rm B}^{(2)})=0\,,~~~~
(Q_{\rm B}^{(2)}+Q_{\rm B}^{(3)})|R(2,3)\rangle=0\,,
\label{eq:BRS_hatgamma_R}
\end{eqnarray}
one can show that 
$(Q_{\rm B}+\bar Q_{\rm B})|V_{\rm c}\rangle=0$ means 
$Q_{\rm B}|\Phi_V\rangle$=0. The other condition (\ref{eq:com_on_closed}),  
required for the gauge invariance of the overlaps, 
can be proven by the generalized gluing and
re-smoothing theorem (\ref{eq:comm_LPP}).
Therefore, if a primary field
$V_{\rm c}(z,\bar z)$ of conformal dimension $(0,0)$ and the ghost
number two is BRST invariant, it yields a gauge invariant overlap 
by $\langle \hat\gamma(1_{\rm c},2)|$ via (\ref{eq:O_Vgamma}).
Therefore, it may be natural to call $\left|\Phi_V\right>$ 
an on-shell closed string state in the open string Hilbert space.
In particular, when $V_{\rm c}(z,\bar z)$ can be put in the form
$V_{\rm c}(z,\bar z)=c(z)\bar c(\bar z)V_{\rm m}(z,\bar z)$,
where $V_{\rm m}(z,\bar z)$ is in the matter sector, 
the BRST invariance is guaranteed, 
if $V_{\rm m}(z,\bar z)$ is a primary field of conformal dimension $(1,1)$. 

For later convenience, it is useful to understand that 
on-shell closed string states are left invariant by the 
transformation generated by $K_n=L_n-(-1)^nL_{-n}$, 
where $L_n$ is the total Virasoro operator with the central charge zero. 
In fact, the requirement that $V_{\rm c}(z,\bar z)$ be a primary field of 
dimension $(0,0)$, $\Phi_V$ means that 
\begin{eqnarray}
 K_n|\Phi_V\rangle=0\,,~~~~~~~(n=1,2,3,\cdots).
\label{eq:symK_n}
\end{eqnarray}
Because $K_n$ is odd under the BPZ conjugation,
the $K_n$ invariance  (\ref{eq:symK_n}) of on-shell closed string states 
means that ${\cal O}_V(\Psi)$ is invariant under the transformation 
$\Psi\to{}e^{\sum_n v_n K_n}\Psi$ on an open string field $\Psi$;
\begin{eqnarray}
 {\cal O}_V(e^{\sum_n v_n K_n}\Psi)= {\cal O}_V(\Psi)
\label{eq:K_nsym}
\end{eqnarray}
for any constants $v_n$. It may be regarded as a part of the gauge symmetry
\cite{IIKTZ, Okawa:2006sn}.
Moreover, if we take a form
$V_{\rm c}(z,\bar z)=c(z)\bar c(\bar z)V_{\rm m}(z,\bar z)$,
where $V_{\rm m}(z,\bar z)$ is a primary field of conformal dimension $(1,1)$,
$\left|\Phi_V\right>$ satisfies
\begin{eqnarray}
&&K_{2m-1}^{({\rm m})}|\Phi_V\rangle=0\,,~~~
(K_{2m}^{({\rm m})}-3(-1)^mm)|\Phi_V\rangle=0\,,
~~~(m=1,2,3,\cdots)
\label{eq:symK_nmat}
\end{eqnarray}
where 
$K_n^{({\rm m})}\equiv  L_n^{({\rm m})}-(-1)^nL_{-n}^{({\rm m})}$ 
for the matter Virasoro operator $L^{({\rm m})}_n$ and also 
\begin{eqnarray}
&&K_{2m-1}^{({\rm gh})}|\Phi_V\rangle=0\,,~~~~~~
(K_{2m}^{({\rm gh})}+3(-1)^mm)|\Phi_V\rangle=0\,,~~~~
~~~(m=1,2,3,\cdots)\,,
\label{eq:symK_ngh}
\end{eqnarray}
where 
$K_n^{({\rm gh})}\equiv  L_n^{({\rm gh})}-(-1)^nL_{-n}^{({\rm gh})}$ 
for the ghost Virasoro operator $L^{({\rm gh})}_n$.
The above relations (\ref{eq:symK_n}), (\ref{eq:symK_nmat})
and (\ref{eq:symK_ngh}) can be derived from more general 
relations given in (\ref{eq:K_nsym_gen_odd}) and
(\ref{eq:K_nsym_gen_even}).

In this paper, we will restrict ourselves to consider 
on-shell closed string states of the form 
\begin{eqnarray}
\left|\Phi_V\right>
&=&\sum_{m,n}\zeta_{mn}c(i)V_m(i)c(-i)V_n(-i)|{\cal I}\rangle,
\label{eq:Phi_V}
\end{eqnarray}
where $\zeta_{mn}$ is the polarization constant with $V_m(z)$ a matter 
primary field on the doubling of the upper half plane.

\section{
Gauge Invariant Overlaps for Schnabl's Solution
\label{sec:GinvSch}}

Let us begin with a brief review on the analytic solution $\Psi_{\lambda}$
discovered by Schnabl in \cite{Schnabl_tach}. 
Schnabl has found the sliver frame very useful to obtain the solution. 
In the frame, a primary field $\tilde\phi(\tilde{z})$ of 
conformal dimension $h$ is related by the conformal transformation 
$\tilde z=\arctan z$ to the usual one $\phi(z)$ on the canonical 
upper half plane as $\tilde \phi(\tilde z)=(1+z^2)^{h}\phi(z)$. 
The field $\tilde\phi(\tilde{z})$ is expanded in terms of 
the oscillators $\tilde\phi_n$ as $\tilde\phi(\tilde{z})=\sum_n
\tilde\phi_n{\tilde{z}}^{-n-h}$. In this paper, 
the tilde $\tilde{\ }$ refers to the sliver frame, as in \cite{Schnabl_tach}. 
In particular, the operators 
$\hat{\cal L}\equiv {\cal L}_0+{\cal L}_0^{\dagger}$, ${\hat {\cal B}}\equiv
{\cal B}_0+{\cal B}_0^{\dagger}$ will be often used in this paper, where 
${\cal L}_0=\tilde L_0$ and ${\cal B}_0=\tilde b_0$
are zero modes of the total energy momentum tensor and $b$-ghost field, 
respectively. 

Making use of the oscillators, Schnabl gave one parameter of solutions 
\begin{eqnarray}
 \Psi_{\lambda}&=&
\sum_{n=0}^{\infty}\sum_{p\ge -1,\,p:{\rm odd}}{(-1)^n\pi^p\over 
n!\,2^{n+2p+1}}f_{n+p+1}(\lambda)\hat{\cal L}^n\tilde{c}_{-p}|0\rangle
\nonumber\\
&&+\sum_{n=0}^{\infty}\sum_{p,q\ge -1,\,p+q:{\rm odd}}
{(-1)^{n+q}\pi^{p+q}\over n!\,2^{n+2(p+q)+3}}f_{n+p+q+2}(\lambda)
\hat{\cal B}\hat{\cal L}^n\tilde{c}_{-p}
\tilde{c}_{-q}|0\rangle\,,
\label{eq:puregauge(I)}
\end{eqnarray}
parametrized by $\lambda$, 
where the function $f_n(\lambda)$ is defined by 
\begin{eqnarray}
f_n(\lambda)&=&
\left\{
\begin{array}[tb]{lc}
B_n&(\lambda=1),
\\
-n\lambda\,{\rm Li}_{1-n}(\lambda)-\delta_{n,1}\lambda
&
(\lambda\ne 1),
\end{array}
\right.
\end{eqnarray}
with the Bernoulli number $B_n$ and the
polylogarithmic function ${\rm Li}_n(z)$. 
The functions $f_n(\lambda)$ can also be organized in the generating 
function 
\begin{eqnarray}
\label{eq:gen_fun}
 \frac{\lambda z}{\lambda
  e^z-1}&=&\sum_{n=0}^{\infty}\frac{f_n(\lambda)}{n!}z^n.
\end{eqnarray}

In order to put the solution (\ref{eq:puregauge(I)}) 
in a somewhat simpler form, it is useful to introduce string fields 
\begin{eqnarray}
 \psi_r&\equiv&{2\over \pi}\hat{U}_{r+2}\!
\left[-\frac{1}{\pi}\hat{\cal B}\tilde{c}(\frac{\pi r}{4})
\tilde{c}(-\frac{\pi r}{4})+\frac{1}{2}(\tilde{c}(-\frac{\pi r}{4})
+\tilde{c}(\frac{\pi r}{4}))
\right]\!|0\rangle
\label{eq:psi_r}
\\
&=&\!\!\underset{p:{\rm odd}}{\sum_{n\ge 0;p\ge -1}}\!
\frac{(-1)^n\pi^p}
{n!2^{n+2p+1}}r^{n+p+1}\hat{\cal L}^n\tilde c_{-p}|0\rangle
+\!\!\underset{p+q:{\rm odd}}{\sum_{n\ge 0;p,q\ge -1}}\!
\frac{(-1)^{n+q}\pi^{p+q}}{n!2^{n+2p+2q+3}}r^{n+p+q+2}
\hat{\cal B}\hat{\cal L}^n\tilde c_{-p}\tilde c_{-q}|0\rangle,\nonumber
\end{eqnarray}
where 
\begin{equation}
\hat{U}_r\equiv U_r^{\dagger}U_r=e^{-\frac{r-2}{2}\hat{\cal L}}
\end{equation}
with $U_r=(2/r)^{{\cal L}_0}$. They can be used to give 
a wedge state $\langle r|=\langle 0|\hat{U}_r=\langle 0|U_r$, 
which is a surface state defined by the conformal map 
$f_r(z)=\tan\left((2/r)\arctan z\right)$.
Using the generating function (\ref{eq:gen_fun}) and the string fields 
$\psi_r$, one can see that the solution $\Psi_{\lambda}$ is given by 
\begin{eqnarray}
\label{eq:sol2}
 \Psi_{\lambda}&=&
\frac{\lambda \partial_r}{\lambda e^{\partial_r}-1}\psi_r|_{r=0}
=\sum_{k=0}^{\infty}\frac{f_k(\lambda)}{k!}\partial_r^k\psi_r|_{r=0}.
\end{eqnarray}

Upon expanding the right hand side of (\ref{eq:sol2}) in terms of 
the derivative $\partial_r$, one can see that it starts with 1 for 
$\lambda=1$, otherwise it starts with $\lambda/(\lambda-1)\partial_r$. 
Therefore, for $\lambda\not=1$, by expanding (\ref{eq:sol2}) formally 
in terms of $\lambda$, one finds that 
\begin{eqnarray}
\label{eq:sol2lambda}
 \Psi_{\lambda\ne 1}&=&
-\lambda\sum_{n=0}^{\infty}\lambda^n 
e^{n\partial_r}\partial_r\psi_r|_{r=0}
=-\sum_{n=0}^{\infty}\lambda^{n+1}\partial_r\psi_r|_{r=n},
\end{eqnarray}
which certainly starts with the first derivative $\partial_r$.
It was in this form (\ref{eq:sol2lambda}) that the solution 
$\Psi_{\lambda\ne 1}$ was shown in \cite{Schnabl_tach} 
to satisfy the equation of motion order by order in $\lambda$.

For $\lambda=1$, exploiting the Euler-Maclaurin formula, 
Schnabl has discussed that the solution has the expansion 
\begin{eqnarray}
 \Psi_{\lambda=1}&=&\lim_{N\to \infty}
\left(\psi_{N+1}-\sum_{n=0}^N\partial_r\psi_r|_{r=n}
\right).
\label{eq:Psi_lambda_1_N}
\end{eqnarray}
The first term $\lim_{N\to\infty}\psi_{N+1}$ is called the phantom term 
and gives finite contributions to the classical action \cite{Schnabl_tach} 
for the solution. The solution wouldn't satisfy the equation of motion 
$\langle \Psi_{\lambda=1},( Q_{\rm B}\Psi_{\lambda=1}
+\Psi_{\lambda=1}*\Psi_{\lambda=1})\rangle=0$ in the `strong' sense 
without the phantom term \cite{Okawa:2006vm, Fuchs:2006hw}. 

In the next two subsections, we will evaluate the gauge invariant 
overlap (\ref{eq:O_VPsi}) for the solution $\Psi_{\lambda}$
in two ways; analytically in the sliver frame 
and numerically with the conventional level truncation.

\subsection{Analytic Evaluation in the Sliver Frame
\label{sec:analytic_S}}

Let us first evaluate the gauge invariant overlap 
${\cal O}_V(\Psi_{\lambda})$ analytically in the sliver frame.
Since one can verify that 
 \begin{eqnarray}
&&U_1^{\dagger}={\rm bpz}(U_{f_{\cal I}})=U^{-1}_{I\circ f_{\cal I}\circ I},
~~~~\phi(z)U_1^{\dagger}=
U_1^{\dagger}(I\circ f_{\cal I}\circ I\circ\phi(z)),\\
&&f_{\cal I}(z)=\frac{2z}{1-z^2}\,,~~~~I(z)=-\frac{1}{z}\,,
\end{eqnarray}
by using them, one finds that  
\begin{eqnarray}
&&\hat U_1^{-1} \phi(e^{i\theta})\hat U_1
=(\cos(it+\frac{\pi}{4}))^{2h}\,\tilde \phi(it)\,,~~~~
e^{i\theta}=\tan(it+\frac{\pi}{4})\,,
\label{eq:hatU_1-1}
\end{eqnarray}
for a primary field $\phi$ of conformal dimension $h$.

Using (\ref{eq:hatU_1-1}), one can rewrite the on-shell closed string state 
(\ref{eq:Phi_V}) as 
\begin{eqnarray}
  \Phi_V&=&\sum_{m,n}\zeta_{mn}\hat{U}_1\tilde c(i\infty)
\tilde V_m(i\infty)\tilde c(-i\infty)
\tilde V_n(-i\infty)|0\rangle
\label{eq:Phi_Vtilde}
\end{eqnarray}
with the operators in the sliver frame, and one may regularize it by 
replacing $\pm i \infty$ by $\pm i M$ in the arguments of the fields 
as 
\begin{eqnarray}
  \Phi_{V,M}&=&\sum_{m,n}\zeta_{mn}\hat{U}_1\tilde c(iM)
\tilde V_m(iM)\tilde c(-iM)
\tilde V_n(-iM)|0\rangle,
\label{eq:Phi_V^M}
\end{eqnarray}
to make well-defined our calculation of the gauge invariant overlap 
(\ref{eq:O_VPsi}).

In order to estimate the gauge invariant overlap (\ref{eq:O_VPsi}) for 
the analytic solution $\Psi_{\lambda}$, 
essentially one needs to calculate the inner product 
$\langle \Phi_{V,M},\psi_r\rangle$. However, instead of it, 
since $\vev{\Phi_{V,M},\psi_r}=\langle{\cal I}|\Phi_{V,M}*\psi_r\rangle
=\langle{\cal I}|\psi_r*\Phi_{V,M}\rangle$, one may compute 
$\langle{\cal I}|\Phi_{V,M}*\psi_r\rangle$ or 
$\langle{\cal I}|\psi_r*\Phi_{V,M}\rangle$.

The sliver frame facilitates the calculation of the star products 
of string fields, and in particular, one finds \cite{Schnabl_tach} that 
\begin{eqnarray}
\hat U_r\tilde{\phi}_1(\tilde{x}_1){\small \cdots} 
\tilde{\phi}_n(\tilde{x}_n)|0\rangle *
\hat U_s\tilde{\psi}_1(\tilde{y}_1){\small \cdots} 
\tilde{\psi}_m(\tilde{y}_m)|0\rangle\nonumber
=\hat U_{r+s-1}
\tilde{\phi}_1(\tilde{x}_1'){\small \cdots} 
\tilde{\phi}_n(\tilde{x}_n')
\tilde{\psi}_1(\tilde{y}_1'){\small \cdots} 
\tilde{\psi}_m(\tilde{y}_m')|0\rangle.
\\
\label{eq:starformula}
\end{eqnarray}
Note that the coordinates on the right hand side are shifted as 
$\tilde x_i'=\tilde x_i+{\pi\over 4}(s-1),\tilde y_j'=\tilde
y_j-{\pi\over 4}(r-1)$. 
Since the operators
\begin{eqnarray}
B^{R}_{1}=\int_{+i\infty-{\pi\over4}}^{-i\infty-{\pi\over4}} 
{d\tilde{z}\over2\pi{i}}\tilde{b}(\tilde{z}),
\qquad
B^{L}_{1}=\int_{-i\infty+{\pi\over4}}^{+i\infty+{\pi\over4}} 
{d\tilde{z}\over2\pi{i}}\tilde{b}(\tilde{z})
\end{eqnarray}
act on the right and the left half-string, they satisfy
\begin{eqnarray}
&&\left(B^{R}_1\Psi_1\right)*\Psi_2=
-(-)^{|\Psi_1|}\Psi_1*\left(B^{L}_1\Psi_2\right),
\label{eq:LRstring}\\
&&B^{L}_1\left(\Psi_1*\Psi_2\right)
=\left(B^{L}_1\Psi_1\right)*\Psi_2,
\qquad
B^{R}_1\left(\Psi_1*\Psi_2\right)
=(-)^{|\Psi_1|}\Psi_1*\left(B^{R}_1\Psi_2\right).
\label{eq:halfstring}
\end{eqnarray}
Therefore, using the above formulae (\ref{eq:LRstring}), 
(\ref{eq:halfstring}) and noting that 
\begin{eqnarray}
B^{R}_{1}={1\over2}B_1-{1\over\pi}\hat{\cal B}, 
\qquad
B^{L}_{1}={1\over2}B_1+{1\over\pi}\hat{\cal B},
\end{eqnarray}
one can verify the formulae 
\begin{eqnarray}
&&(\hat{\cal B}\Psi_1)*\Psi_2=\hat{\cal
 B}(\Psi_1*\Psi_2)+(-1)^{|\Psi_1|}\frac{\pi}{2}
\Psi_1*B_1\Psi_2\,,
\label{eq:B_1form}\\
&&\Psi_1*(\hat{\cal B}\Psi_2)=(-1)^{|\Psi_1|}\hat{\cal
 B}(\Psi_1*\Psi_2)-(-1)^{|\Psi_1|}\frac{\pi}{2}
(B_1\Psi_1)*\Psi_2\,,
\label{eq:B_2form}
\end{eqnarray}
where $B_1=\tilde b_{-1}=b_1+b_{-1}$.
Using them and the formula (\ref{eq:starformula}), 
one finds that the star products $\Phi_{V,M}*\psi_r$ and $\psi_r*\Phi_{V,M}$ 
are given by 
\begin{eqnarray}
&&\Phi_{V,M}*\psi_r=\frac{1}{\pi}\sum_{m,n}\zeta_{mn}\hat U_{r+2}\biggl(
\!\tilde c\tilde V_m(iM+\frac{\pi}{4}(r+1))
\tilde c\tilde V_n(-iM+\frac{\pi}{4}(r+1))\!
\left(\tilde c(-\frac{\pi}{4}r)+\tilde c(\frac{\pi}{4}r)\right)
\nonumber\\
&&~~~~~~~~~~~~~~~+\bigl(\tilde V_m(iM+\frac{\pi}{4}(r+1))
\tilde c\tilde V_n(-iM+\frac{\pi}{4}(r+1))
\label{eq:Phi_psir}\\
&&~~~~~~~~~~~~~~~-\tilde c\tilde V_m(iM+\frac{\pi}{4}(r+1))
\tilde V_n(-iM+\frac{\pi}{4}(r+1))
\bigr)\tilde c(\frac{\pi}{4}r)\tilde c(-\frac{\pi}{4}r)
\nonumber\\
&&~~~~~~~~~~~~~~~-\frac{2}{\pi}\hat{\cal B}
\tilde c\tilde V_m(iM+\frac{\pi}{4}(r+1))
\tilde c\tilde V_n(-iM+\frac{\pi}{4}(r+1))
\tilde c(\frac{\pi}{4}r)\tilde c(-\frac{\pi}{4}r)
\biggr)|0\rangle,\nonumber
\end{eqnarray}
and
\begin{eqnarray}
&&\psi_r*\Phi_{V,M}=\frac{1}{\pi}\sum_{m,n}\zeta_{mn}
\hat U_{r+2}\biggl(\!
\tilde c\tilde V_m(iM-\frac{\pi}{4}(r+1))
\tilde c\tilde V_m(-iM-\frac{\pi}{4}(r+1))\!
\left(\tilde c(-\frac{\pi}{4}r)+\tilde c(\frac{\pi}{4}r)\right)
\nonumber\\
&&~~~~~~~~~~~~~~~-\bigl(\tilde V_m(iM-\frac{\pi}{4}(r+1))
\tilde c\tilde V_m(-iM-\frac{\pi}{4}(r+1))
\label{eq:psir_Phi}\\
&&~~~~~~~~~~~~~~~-\tilde c\tilde V_m(iM-\frac{\pi}{4}(r+1))
\tilde V_m(-iM-\frac{\pi}{4}(r+1))
\bigr)\tilde c(\frac{\pi}{4}r)\tilde c(-\frac{\pi}{4}r)
\nonumber\\
&&~~~~~~~~~~~~~~~-\frac{2}{\pi}\hat{\cal B}
\tilde c\tilde V_m(iM-\frac{\pi}{4}(r+1))
\tilde c\tilde V_m(-iM-\frac{\pi}{4}(r+1))
\tilde c(\frac{\pi}{4}r)\tilde c(-\frac{\pi}{4}r)
\biggr)|0\rangle,
\nonumber
\end{eqnarray}
respectively.

Furthermore, substituting the above results 
into $\langle{\cal I}|\Phi_{V,M}*\psi_r\rangle$ and 
$\langle{\cal I}|\psi_r*\Phi_{V,M}\rangle$ and using the relations 
\begin{eqnarray}
&&\langle {\cal I}|\hat{U}_{r+2}=\langle 0|U_{r+1}\,,~~~
\langle {\cal I}|\hat{U}_{r+2}\hat{\cal B}=
\frac{2}{r+1}\langle 0|U_{r+1}{\cal B}_0,
\end{eqnarray}
one finds that $\langle{\cal I}|\Phi_{V,M}*\psi_r\rangle$ and 
$\langle{\cal I}|\psi_r*\Phi_{V,M}\rangle$ give the same results 
\begin{eqnarray}
&&\langle \Phi_{V,M},\psi_r\rangle=
\langle {\cal I}|\Phi_{V,M}*\psi_r\rangle
=\langle {\cal I}|\psi_r*\Phi_{V,M}\rangle
\nonumber\\
&&=\frac{1}{2\pi}\sum_{m,n}\zeta_{mn}\,
{}_{\rm mat}\langle 0|\tilde V_m(\frac{2iM}{r+1}\pm \frac{\pi}{2})\tilde
V_m(\frac{-2iM}{r+1}\pm \frac{\pi}{2})|0\rangle_{\rm mat}
\nonumber\\
&&~~\times \biggl(
-\frac{4iM}{\pi}{}_{bc}\langle 0|
\left(\tilde c(\frac{-2iM}{r+1}\pm \frac{\pi}{2})
+\tilde c(\frac{2iM}{r+1}\pm \frac{\pi}{2})
\right)\tilde c(\frac{\pi r}{2(r+1)})
\tilde c(\frac{-\pi r}{2(r+1)})|0\rangle_{bc}
\nonumber\\
&&~~~~~~~+{}_{bc}\langle 0|
\tilde c(\frac{2iM}{r+1}\pm \frac{\pi}{2})
\tilde c(\frac{-2iM}{r+1}\pm \frac{\pi}{2})
\left(\tilde c(\frac{-\pi r}{2(r+1)})+
\tilde c(\frac{\pi r}{2(r+1)})\right)|0\rangle_{bc}
\biggr)
\nonumber\\
&&=\frac{C_V}{2\pi i}
\left(
\sinh\frac{4 M}{r+1}-\frac{4 M}{\pi}\sin\frac{\pi}{r+1}
\right)
\left(
\cosh\frac{4 M}{r+1}-\cos\frac{\pi}{r+1}
\right)
\left(\sinh \frac{4 M}{r+1}\right)^{-2},
\nonumber\\
\label{eq:naiseki}
\end{eqnarray}
where the factor $C_V$ is given by
\begin{eqnarray}
 C_V&=&{}_{\rm mat}\langle 0|0\rangle_{\rm mat}
\sum_{m,n}\zeta_{mn}v_{mn}.
\label{eq:C_V_form}
\end{eqnarray}
The constants $v_{mn}$ are the metric appearing 
in the OPE of the matter primary fields $V_m(z)$ of 
conformal dimension one as 
\begin{eqnarray}
\label{eq:VVope}
 V_m(y)V_n(z)&\sim&\frac{v_{mn}}{(y-z)^2}+{\rm finite}\,.~~~~
~(y\to z)
\end{eqnarray}
By taking the limit $M\to +\infty$ in (\ref{eq:naiseki}), 
one obtains 
\begin{eqnarray}
 \langle \Phi_V,\psi_r\rangle=\lim_{M\to +\infty}\langle
  \Phi_{V,M},\psi_r\rangle=\frac{C_V}{2\pi i}\,.
\end{eqnarray}
It is interesting to note that $\langle \Phi_V,\psi_r\rangle$ is 
independent of $r$, and thus one can see that 
$\partial_r\langle \Phi_V,\psi_r\rangle=0$. 
It in turn means that 
\begin{eqnarray}
{\cal O}_V(\Psi_{\lambda})&=&
\sum_{k=0}^{\infty}\frac{f_k(\lambda)}{k!}\partial_r^k
\langle \Phi_V,\psi_r\rangle|_{r=0}
=f_0(\lambda)\langle \Phi_V,\psi_0\rangle
\nonumber\\
&=&
\left\{
\renewcommand{\arraystretch}{1.8}
\begin{array}[tb]{lc}
\displaystyle  \frac{C_V}{2\pi i}& (\lambda=1),\\
\displaystyle 0&(\lambda\ne 1).
\end{array}
\renewcommand{\arraystretch}{1}
\right.
\label{eq:O_VPsi_l_a}
\end{eqnarray}
Namely, the value can be nonzero only for $\lambda=1$.
It seems that this nonzero value only comes from 
the inner product with the phantom term $\psi_{N+1}$, 
if one uses the expression given in (\ref{eq:Psi_lambda_1_N}).
From this viewpoint, we should not take the limit $N\to \infty$ first because
the inner product vanishes for finite $M$ in (\ref{eq:naiseki}); 
$\lim_{N\to \infty}\langle \Phi_{V,M},\psi_{N+1}\rangle=0$, which 
means that the order of the two limits isn't interchangeable 
\begin{eqnarray}
 \lim_{N\to \infty}\biggl(\lim_{M\to +\infty}\langle 
  \Phi_{V,M},\psi_{N+1}\rangle\biggr)&\ne&
\lim_{M\to +\infty}\biggl(\lim_{N\to \infty}\langle
  \Phi_{V,M},\psi_{N+1}\rangle\biggr)\,.
\end{eqnarray}
The order on the left hand side is consistent with our evaluation with  
the expression (\ref{eq:sol2}) to yield the result (\ref{eq:O_VPsi_l_a}).
Rearranging the terms in the expression (\ref{eq:Psi_lambda_1_N}), 
one obtains 
\begin{eqnarray}
 \Psi_{\lambda=1}
&=&\psi_0
+\sum_{n=0}^{\infty}\left(\psi_{n+1}-\psi_n-\partial_r\psi_r|_{r=n}\right),
\end{eqnarray}
and there seems no problem with the ordering of the limits. 

Anyhow, it is obvious that there are subtleties with the regularization 
with the cutoff $M$ and the order of the limits. In order to confirm 
our results in this subsection, we will evaluate the gauge invariant 
overlap numerically in the level truncation calculation. 

\subsection{Numerical Evaluation with the Level Truncation
\label{sec:numerical_Sch}}

In this section, the gauge invariant product for the solution $\Psi_\lambda$ 
will be calculated by the level truncation calculation to confirm 
our analytic results in the previous subsection, which is 
a similar strategy to the calculation for the vacuum energy 
for the solution \cite{Schnabl_tach,Takahashi:2007du}.

As usual, we begin with expanding a string field $\Psi$ in terms of 
the Fock space states as 
\begin{eqnarray}
  \Psi= t\,c_1\ket{0}
                +u\,c_{-1}\ket{0}
                +v\,(\alpha_{-1}\cdot\alpha_{-1})c_1\ket{0}
                +w\,b_{-2}c_0c_1\ket{0}+\cdots,
\end{eqnarray}
where the dots $\cdots$ denote higher level terms than level 2. 
Substituting it into (\ref{eq:O_VPsi}), one obtains 
the gauge invariant overlap ${\cal O}_V(\Psi)$ in term of 
the component fields $t,u,v,\cdots$.

For example, let us consider the on-shell closed string tachyon state 
(\ref{eq:closed_tach}) 
in the open string field theory on a D$p$-brane in the flat 26-dimensional 
spacetime. For simplicity, we set the momentum of the string field along 
the Neumann directions to zero. Using the oscillator expression
in appendix \ref{sec:Ginv_closed}, the gauge invariant overlap for the
tachyon state gives\footnote{In order for the overlap
(\ref{eq:Zwiebach_inv}) to be nonzero, 
one needs to consider open string field theory on a D-brane with 
at least one Dirichlet direction, due to the momentum conservation.}
\begin{eqnarray}
\label{eq:ginv_level2}
 {\cal O}_k(\Psi)=\frac{1}{4}t-\frac{3}{2}v+\frac{1}{4}u+\cdots.
\end{eqnarray}
Similarly, the gauge invariant overlap for the on-shell closed string 
dilaton state 
can be calculated by using the oscillator expression 
in appendix \ref{sec:Ginv_closed}.

In the level truncation calculation, one first takes the limit $N\to\infty$ 
in (\ref{eq:Psi_lambda_1_N}), while keeping the level fixed. 
In this limit, as pointed out in \cite{Schnabl_tach}, the phantom term in 
(\ref{eq:Psi_lambda_1_N}) can be neglected in the solution $\Psi_\lambda$, 
which thus allows one to treat (\ref{eq:sol2lambda}) and 
(\ref{eq:Psi_lambda_1_N}) without any distinction as 
\begin{eqnarray}
 \Psi_\lambda =-\sum_{n=0}^\infty\lambda^{n+1}\left.
\partial_r\psi_r\right|_{r=n}.
\end{eqnarray}
Expanding $\psi_r$ in terms of the oscillators, as given in appendix
\ref{sec:leveltr}, 
substituting them into the analytic solution $\Psi_\lambda$, one obtains 
the values of the component fields $t,u,v,\cdots$. In fact, up to the level 
2, one finds that 
\begin{eqnarray}
 t &=& \sum_{n=2}^\infty
       \lambda^{n-1}\frac{d}{dn}
       \left[\frac{n}{\pi}\sin^2\left(\frac{\pi}{n}\right)
       \left(-1+\frac{n}{2\pi}\sin\left(\frac{2\pi}{n}\right)
       \right)\right],\\
 u &=& \sum_{n=2}^\infty
       \lambda^{n-1}\frac{d}{dn}
       \left[\left(\frac{4}{n\pi}-\frac{n}{\pi}\sin^2\left(
       \frac{\pi}{n}\right)\right)\left(
       -1+\frac{n}{2\pi}\sin\left(\frac{2\pi}{n}\right)
       \right)\right],\\
 v &=& \sum_{n=2}^\infty
       \lambda^{n-1}\frac{d}{dn}
       \left[\left(\frac{4}{3n\pi}-\frac{n}{3\pi}\right)
       \sin^2\left(\frac{\pi}{n}\right)\left(
       -1+\frac{n}{2\pi}\sin\left(\frac{2\pi}{n}\right)
       \right)\right],\\
 w &=& \sum_{n=2}^\infty
       \lambda^{n-1}\frac{d}{dn}
       \left[\sin^2\left(\frac{\pi}{n}\right)\left(
       \frac{8}{3n\pi}-\frac{2n}{3\pi}+\frac{n^2}{3\pi^2}
       \sin\left(\frac{2\pi}{n}\right)\right)\right].
\end{eqnarray}
The series in $\lambda$ 
can be evaluated numerically for $-1\leq \lambda\leq 1$ with arbitrary
precision \cite{Schnabl_tach,Takahashi:2007du},
although it seems formidable to calculate them analytically.

Evaluating the above infinite sums numerically and substituting them 
into (\ref{eq:ginv_level2}), one obtains the gauge invariant overlap 
for the analytic solution up to the level 2, which is plotted out on 
Figure 3. The resulting numerical value $0.149\cdots$ for $\lambda=1$ 
is about 94\% of $1/2\pi$ and thus is very close to the analytical
result (\ref{eq:O_k_analytic}), as will be seen soon. 
\begin{figure}[h]
\centerline{\includegraphics[width=10cm]{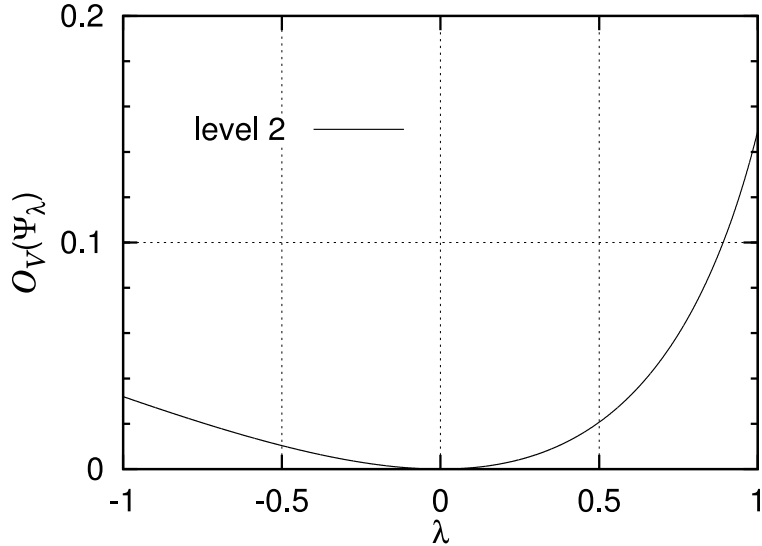}}
\caption{The gauge invariant overlap for the analytic solution 
evaluated by the level 2 truncation in the closed tachyon and dilaton cases.}
\label{fig:level2}
\end{figure}

Let us further move on higher level calculations of the gauge
invariant overlap up to the level 14. 
The analytic solution $\Psi_{\lambda}$ is given by the sum of 
the direct products of the matter Fock space and the ghost one, 
the former of which may be written in terms of the matter Virasoro 
operators $L^{({\rm m})}_n$ acting on the $SL(2,{\bf R})$ invariant
vacuum $\ket{0}$. 
Furthermore, as discussed in appendix \ref{sec:tachyon=dilaton}, one can
rewrite it 
in terms of the operators $K^{({\rm m})}_n$ acting on the vacuum $\ket{0}$, 
instead of $L^{({\rm m})}_n$. Although the total operator
$K_n=K^{({\rm m})}_n+K^{({\rm gh})}_n$ 
annihilates the on-shell closed string states $\ket{\Phi_V}$, 
the states $\ket{\Phi_V}$ are the eigenstates of the matter part
$K^{({\rm m})}_n$. 
Since the ghost part of $\ket{\Phi_V}$ doesn't depend on which closed string 
state one chooses for $\ket{\Phi_V}$, the eigenvalues of the matter part 
$K^{({\rm m})}_n$ for the eigenstates $\ket{\Phi_V}$ are all the same
and 
are given in (\ref{eq:symK_nmat}). This fact is interesting and indeed
facilitates higher level calculations of the gauge invariant overlap up
to the level 14.

More precisely, let us suppose to take the level truncated solution 
up to the level $L$ 
\begin{eqnarray}
 \Psi_{\lambda,L}&=&
-\sum_{0\le 2l\le L}
\sum_{n=2}^{\infty}\lambda^{n-1}\partial_r\psi_{r-2,2l}|_{r=n},
\end{eqnarray}
where the additional suffix $2l$ on $\psi_r$ denotes the level.
The state $\psi_{r-2,2l}$ is given by the products of the matter and ghost 
sectors as 
\begin{eqnarray}
 \psi_{r-2,2l}&=&
\sum_{j=0}^l\psi_{r-2,2j}^{({\rm m})}\otimes \psi_{r-2,2l-2j}^{({\rm
gh})}.
\label{eq:psir-2L}
\end{eqnarray}
It follows from the discussion in appendix \ref{sec:leveltr} that 
$\psi_{r-2,L}^{({\rm m})}$ and $\psi_{r-2,L}^{({\rm gh})}$ 
can be read by taking the terms of the level $L$ from 
\begin{eqnarray}
\label{eq:psimr}
 |\psi_{r-2}^{({\rm m})}\rangle
&=&\cdots e^{u_{6}(r)L_{-6}^{({\rm m})}}
e^{u_{4}(r)L_{-4}^{({\rm m})}}e^{u_{2}(r)L_{-2}^{({\rm
m})}}|0\rangle_{\rm mat},\\
\label{eq:psighr}
 |\psi_{r-2}^{({\rm gh})}\rangle
&=&\cdots e^{u_{6}(r)L_{-6}^{({\rm gh})}}
e^{u_{4}(r)L_{-4}^{({\rm gh})}}e^{u_{2}(r)L_{-2}^{({\rm
gh})}}(2/r)^{L_0^{({\rm gh})}}|\chi_{r-2}\rangle_{\rm gh},
\end{eqnarray}
where $u_{2k}(r)$ and $|\chi_{r-2}\rangle_{\rm gh}$ are defined in 
(\ref{eq:u2_formula}), (\ref{eq:u2k_formula}) and (\ref{eq:chi_r-2}).
One can see from (\ref{eq:psi_r-2_level}) that $\psi_{r-2,L}^{({\rm m})}$ and 
$\psi_{r-2,L}^{({\rm gh})}$ are zero for odd $L$.

Substituting (\ref{eq:psir-2L}) into the gauge invariant overlap 
with a closed string state $\Phi_V$, one obtains 
\begin{eqnarray}
 {\cal O}_{V}(\Psi_{\lambda,L})&=&
-\sum_{0\le 2l\le L}\sum_{n=2}^{\infty}\lambda^{n-1}\partial_r
G_{2l}(r)|_{r=n}\,,
\label{eq:sum_G_l}
\end{eqnarray}
with 
\begin{eqnarray}
G_{2l}(r)&=&\sum_{j=0}^l
\langle \Phi_{V}^{({\rm m})}|\psi_{r-2,2j}^{(\rm m)}\rangle\cdot
\langle 0|
c_{-1}c_0e^{-\sum_{n=1}^{\infty}(-1)^nc_nb_n}|\psi_{r-2,2l-2j}^{(\rm
gh)}\rangle\,.
\end{eqnarray}
Note that the ghost matrix elements in $G_{2l}(r)$ don't depend on the 
closed string state $\Phi_V$, as mentioned above.
To evaluate the matter matrix elements, the important
point is that the state $\psi_{r-2,2j}^{(\rm m)}$ is expressed only by
using the matter Virasoro operators acting on the vacuum.
From (\ref{eq:psimr}), one can see that
\begin{eqnarray}
 |\psi_{r-2,2j}^{(\rm m)}\rangle
=|0\rangle_{\rm mat}
+u_2(r)L_{-2}^{(\rm m)}|0\rangle_{\rm mat}
+u_4(r)L_{-4}^{(\rm m)}|0\rangle_{\rm mat}
+\frac{1}{2!}u_2(r)^2(L_{-2}^{(\rm m)})^2
|0\rangle_{\rm mat}+\cdots.
\end{eqnarray}
As discussed in detail in appendix \ref{sec:tachyon=dilaton}, one can
give the Fock state 
$L_{-n_1}\cdots{L}_{-n_p}\ket{0}$ as a linear combination of the 
Fock states $K_{m_1}\cdots{K}_{m_r}\ket{0}$. Therefore, one finds 
that $|\psi_{r-2,2j}^{(\rm m)}\rangle$ can be given in terms of them as
\begin{eqnarray}
  |\psi_{r-2,2j}^{(\rm m)}\rangle
&=&\left(1+\frac{13}{2!}u_2(r)^2\right)|0\rangle_{\rm mat}
-u_2(r)K_{2}^{(\rm m)}|0\rangle_{\rm mat}
-u_4(r)K_{4}^{(\rm m)}|0\rangle_{\rm mat}\nonumber\\
&&+\frac{1}{2!}u_2(r)^2(K_{2}^{(\rm m)})^2
|0\rangle_{\rm mat}+\cdots.
\end{eqnarray}
Since $\langle\Phi_V^{(\rm m)}|$ is an
eigenstate of $K_n^{(\rm m)}$ as shown in (\ref{eq:symK_nmat}),
one can immediately obtain the matter matrix
elements in $G_{2l}(r)$ up to the normalization.
The normalization factor is determined by $\zeta_{mn}$ in the closed
string state (\ref{eq:Phi_V}) and $v_{mn}$ in
the OPE of $V_m(z)$ in (\ref{eq:VVope}).

\begin{figure}[htbp]
\begin{center}
\includegraphics[width=13cm]{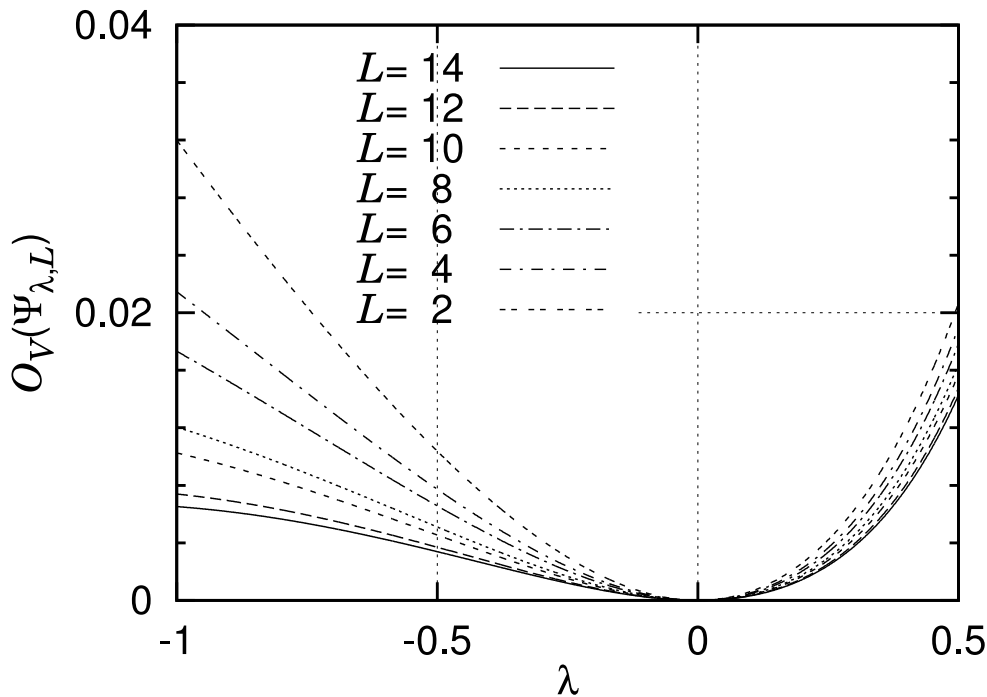}
\end{center}
\caption{Plots of ${\cal O}_{V}(\Psi_{\lambda,L})$
for $-1\leq \lambda\leq 0.5$ at the levels $L=2,4,6,8,10,12,14$}
\label{fig:upto14}
\end{figure}

\begin{figure}[htbp]
\begin{center}
\includegraphics[width=14cm]{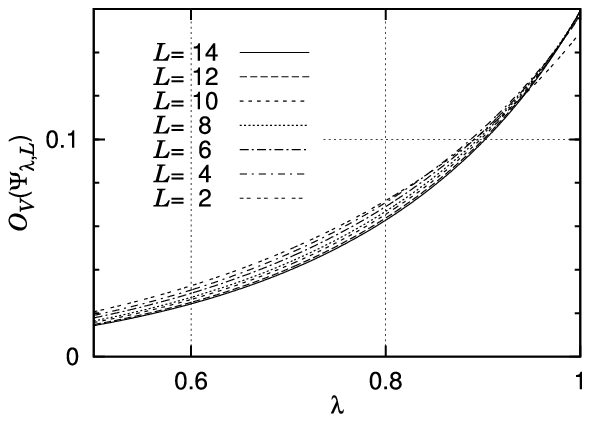}
\end{center}
\caption{Plots of ${\cal O}_{V}(\Psi_{\lambda,L})$
for $0.5\leq \lambda \leq 1$ at the levels $L=2,4,6,8,10,12,14$.}
\label{fig:upto14-2}
\end{figure}

Finally, we have only to evaluate the infinite series (\ref{eq:sum_G_l})
numerically, and one finds that 
\begin{equation}
a^{(L)}_{n}/a^{(L)}_{n+1}=1+4/n+O(1/n^2),
\end{equation} 
where $a_n^{(L)}\equiv-\sum_{0\le 2l\le L}\partial_r G_{2l}(r)|_{r=n}$
for $L=0,\cdots,14$. Therefore, one can see that this series converges 
absolutely for $-1\le \lambda\le 1$. 
With our normalization ${}_{\rm mat}\langle 0|\Phi_V^{({\rm {m}})}\rangle=1/4$,
the numerical results of ${\cal O}_{V}(\Psi_{\lambda,L})$ are depicted in
Figures~\ref{fig:upto14} and \ref{fig:upto14-2}.


This normalization is equivalent to $C_V=i$ in
(\ref{eq:O_VPsi_l_a}), and thus the corresponding analytic result
yields 
\begin{eqnarray}
{\cal O}_{V}(\Psi_{\lambda})&=&\left\{
\renewcommand{\arraystretch}{1.8}
\begin{array}[tb]{cl}
 \displaystyle \frac{1}{2\pi}& (\lambda=1)\\
0&(-1\leq \lambda< 1).
\end{array}
\renewcommand{\arraystretch}{1}
\right.\,
\label{eq:O_k_analytic}
\end{eqnarray}
From Figures 4 and 5, one can observe 
that the plots of
${\cal O}_{V}(\Psi_{\lambda,L})$ ($-1\le \lambda\le 1$)
approach to the analytical result as the truncation level $L$ 
increases. As seen in Figure~\ref{fig:upto14-2}, the resulting plots
around $\lambda\sim 1$ are getting close to the analytic result 
while oscillating. 
In particular, at $\lambda=1$, it approaches to the analytic value
$1/(2\pi)\simeq 0.1591549$, as can seen in Table 1. 
The numerical result for $L=14$ is in remarkably good agreement with 
the analytic one. 
\begin{table}[h]
\begin{center}
\begin{tabular}{|c|c|c|c|c|c|c|c|}
\hline
 $L=0$ & $L=2$ & $L=4$ & $L=6$ & $L=8$ & $L=10$ & $L=12$ & $L=14$\\
\hline
0.138366 &0.149284 &0.156857 &0.157395 & 0.158795 & 0.158765 
 & 0.15922 & 0.159159\\
\hline
\end{tabular}
\end{center}
\caption{The gauge invariant overlap ${\cal O}_{V}(\Psi_{\lambda,L})$
at $\lambda=1$ for various truncation levels of the analytic solution.}
\end{table}

The results on  the gauge invariant overlap ${\cal O}_{V}(\Psi)$
for the solution $\Psi_{\lambda}$ give another evidence 
that $\Psi_{\lambda=1}$ is a nontrivial solution 
and that $\Psi_{\lambda}$ ($-1\le \lambda < 1$) can be gauged away to be 
trivial, and it is also consistent with the result of the vacuum energy 
of the solution $\Psi_{\lambda}$ 
\begin{eqnarray}
S[\Psi_{\lambda}]&=&\left\{
\renewcommand{\arraystretch}{1.8}
\begin{array}[tb]{cl}
\displaystyle
 \frac{1}{2\pi^2g^2}& (\lambda=1),\\
0&(-1\le \lambda < 1),
\end{array}
\renewcommand{\arraystretch}{1}
\right.
\end{eqnarray}
obtained analytically as well as numerically 
in \cite{Schnabl_tach,Okawa:2006vm, Fuchs:2006hw, Takahashi:2007du}.

\section{
Gauge Invariant Overlaps for the Numerical Solution in
Siegel Gauge
\label{sec:Siegel}}

A numerical solution for tachyon condensation was initially obtained by 
Sen and Zwiebach with the level truncation in the Siegel gauge 
\cite{Sen:1999nx} and was improved by going to higher levels 
\cite{Moeller:2000xv,Gaiotto:2002wy}.
One may thus suspect that it can be gauge equivalent to the level
truncated form of the exact solution in \cite{Schnabl_tach}. 
In this section, the gauge invariant overlap for a numerical
solution $\Psi_{\rm N}$ given by \cite{Sen:1999nx,Moeller:2000xv,Gaiotto:2002wy}
will be calculated to compare with the results for Schnabl's solution 
$\Psi_{\lambda}$ in the previous section. 

In order to obtain the numerical solution $\Psi_{{\rm N}}$ in 
the level $(L,2L)$ and $(L,3L)$ approximations, 
following \cite{Sen:1999nx,Moeller:2000xv,Gaiotto:2002wy}, 
one needs to expand an open string field in terms of the usual Fock 
states and truncate it up to the $L_0$-level $L$ with 
the assumption that it is Lorentz scalar and twist even with 
zero momentum. Substituting it into the action and keeping 
the interaction terms up to the total $L_0$-level $2L$ and $3L$ 
in the level $(L,2L)$ and $(L,3L)$ approximations, 
respectively, one obtains the resulting actions for the truncated string 
fields, and one can find the stationary points $\Psi_{{\rm N},(L,2L)}$ and 
$\Psi_{{\rm N},(L,3L)}$ of them in the level $(L,2L)$ and $(L,3L)$ 
approximations, respectively.
In fact, we have confirmed that the vacuum energies of them are in 
agreement with the previous results in 
\cite{Sen:1999nx,Moeller:2000xv,Gaiotto:2002wy}.

Let us now consider the gauge invariant overlaps for the numerical 
solutions $\Psi_{{\rm N},(L,2L)}$ and 
$\Psi_{{\rm N},(L,3L)}$ with the on-shell closed string tachyon 
and dilaton states. 
In the Siegel gauge, an open string field can be expanded up to the 
level 4 as
\begin{eqnarray}
 \Psi&=&
t_0\,c_1\ket{0}
+t_1\,(\alpha_{-1}\cdot \alpha_{-1})c_1\ket{0}
+t_2\,b_{-1}c_{-1}c_1\ket{0}
+t_3\,(\alpha_{-1}\cdot \alpha_{-1})(\alpha_{-1}\cdot \alpha_{-1})
 c_1\ket{0}\nonumber\\
&&
+t_4\,(\alpha_{-2}\cdot \alpha_{-2})c_1\ket{0}
+t_5\,(\alpha_{-1}\cdot \alpha_{-3})c_1\ket{0}
+t_6\,(\alpha_{-1}\cdot \alpha_{-1})b_{-1}c_{-1}c_1\ket{0}\nonumber\\
&&+t_7\,b_{-1}c_{-3}c_1\ket{0}
+t_8\,b_{-2}c_{-2}c_1\ket{0}
+t_9\,b_{-3}c_{-1}c_1\ket{0}+\cdots,
\end{eqnarray}
where $t_i$ denotes component fields.
Using the oscillator expressions of the on-shell closed string states 
in appendix \ref{sec:Ginv_closed}, one obtains 
the gauge invariant overlap for the tachyon as
\begin{eqnarray}
\label{eq:ginv-tachyon}
{\cal O}_k(\Psi)=
\frac{1}{4}t_0 - \frac{3}{2}\,t_1 
-\frac{1}{4}\,t_2 - 10\,t_3 - 13\,t_4 + 8\,t_5 + \frac{3}{2}\,t_6 + 
\frac{1}{4}\,t_8
+\cdots,
\end{eqnarray}
and for the dilaton as
\begin{eqnarray}
\label{eq:ginv-dilaton}
{\cal O}_\eta(\Psi)=
\frac{1}{4}\,t_0 - \frac{3}{2}\,t_1 - \frac{1}{4}t_2 
- 266\,t_3 - 141\,t_4 + 72\,t_5 
+ \frac{3}{2}\,t_6 + \frac{1}{4}\,t_8+\cdots.
\end{eqnarray}
Plugging the stationary points $\Psi_{{\rm N},(L,2L)}$ and 
$\Psi_{{\rm N},(L,3L)}$ for the component fields $t_i$ into 
(\ref{eq:ginv-tachyon}) and (\ref{eq:ginv-dilaton}), one finds 
the results in Table 2.
Note that the results for the tachyon should be identical to the one 
for the dilaton at each of the level approximations, 
with the proper normalization for the closed string states
in appendix \ref{sec:tachyon=dilaton}. This has the same reason as what
was already 
discussed in the previous section \ref{sec:numerical_Sch}. In fact, 
the numerical solution $\Psi_{\rm N}$ can also be given by a linear 
combination of states which are the vacuum $\ket{0}$ on which only 
the matter Virasoro operators $L_n^{({\rm m})}$ and the ghost oscillators act. 
The matter Virasoro operators $L^{({\rm m})}_n$ acting on $\ket{0}$ can
be given 
by a linear combination of the operators $K^{({\rm m})}_n$ acting on
$\ket{0}$. 
The on-shell closed string states are eigenstates of $K^{({\rm m})}_n$ and 
have the same eigenvalue. Furthermore, their ghost parts don't depend on 
which closed string state one chooses. Therefore, up to the overall 
normalization, their gauge invariant overlaps should be the same. 
\begin{table}[h]
\begin{center}
\begin{tabular}[tb]{|c||c|c|c|c|c|c|}
\hline
 $L$ & $L=0$ & $L=2$ & $L=4$& $L=6$ & $L=8$ & $L=10$ \\
\hline
\!${\cal O}_{k/\eta}(\Psi_{{\rm N},(L,2L)})$\!
&0.114044 &0.139790 &0.147931 &0.151225 & 0.152887 & 0.154029 \\
\hline
\!${\cal O}_{k/\eta}(\Psi_{{\rm N},(L,3L)})$\!
&0.114044 &0.141626 &0.148325 &0.151369 & 0.152976 & 0.154080\\
\hline
\end{tabular}
\end{center}
\caption{The gauge invariant overlap ${\cal O}_{k/\eta}(\Psi_{\rm N})$
for the numerical tachyon vacuum solution in the Siegel gauge.}
\end{table}

One can see from Table 2 that 
the value of the gauge invariant overlaps approaches to $1/(2\pi)$ 
as the level increases, as in the case of the Schnabl's solution 
$\Psi_{\lambda=1}$. The best approximation $(10,30)$ for the overlap gives 
97\% of $1/(2\pi)$. In addition to the matching of the D-brane tension, 
this result gives another evidence for the gauge equivalence of 
$\Psi_{\rm N}$ with $\Psi_{\lambda=1}$.

\section{
Gauge Invariant Overlaps for the Marginal Solution
\label{sec:GinvMarg}}

As alternative exact classical solutions to the equation of motion of open 
string field theory, marginal solutions for nonsingular currents 
\cite{Schnabl:2007az, Kiermaier:2007ba, Kishimoto:2007bb} are known 
(for marginal solutions with singular currents, see 
\cite{Takahashi:2001pp,Fuchs:2007yy,Kiermaier:2007vu}),
and it turns out that the gauge invariant overlaps for it can 
explicitly be evaluated similarly to \S\ref{sec:analytic_S}.
In this section, it will indeed be done.

Let us consider the marginal solution 
$\Psi^{(\alpha,\beta)}(\lambda_{\rm m}\hat\psi_{\rm m})$, 
which can be obtained from 
the BRST invariant and nilpotent string field $\hat\psi_{\rm m}=
\hat U_1\tilde c\tilde J(0)|0\rangle$
with a nonsingular marginal current $J$ \cite{Kishimoto:2007bb} as
\begin{eqnarray}
&&\Psi^{(\alpha,\beta)}(\lambda_{\rm
m}\hat\psi_{\rm m})=P_{\alpha}*\frac{1}{1+
\lambda_{\rm m}\hat\psi_{\rm m}*A^{(\alpha+\beta)}}*
\lambda_{\rm m}\hat\psi_{\rm m}*P_{\beta}
=\sum_{n=1}^{\infty}\lambda_{\rm m}^n\psi_{{\rm m},n}\,,
\label{eq:marginal_ab}
\end{eqnarray}
where $P_{\alpha}=\hat{U}_{\alpha+1}|0\rangle,~
A^{(\gamma)}=\frac{\pi}{2}\int_0^{\gamma}d\alpha
B_1^LP_{\alpha}$, and each term on the right hand side yields 
\begin{eqnarray}
\psi_{{\rm m},1}&=&\hat U_{\alpha+\beta+1}
\tilde c\tilde J\biggl(\frac{\pi}{4}(\beta-\alpha)\biggr)|0\rangle\,,\\
\psi_{{\rm
 m},k+1}
&=&\left(-\frac{\pi}{2}\right)^k\int_0^{\alpha+\beta}dr_1
\cdots\int_0^{\alpha+\beta}dr_k\,
\hat U_{\gamma^{(k)}+1}
\prod_{m=0}^k\tilde J(\tilde x_m^{(k)})\nonumber\\
&&\quad\times \biggl[-\frac{1}{\pi}\hat{\cal B}
\tilde c(\tilde x_0^{(k)})\tilde c(\tilde x_k^{(k)})
+\frac{1}{2}\left(\tilde c(\tilde x_0^{(k)})+
\tilde c(\tilde x_k^{(k)})\right)
\biggr]|0\rangle\,,
\end{eqnarray}
where the constants $\gamma^{(k)}$ and the arguments $\tilde x_m^{(k)}$ 
are given by 
\begin{eqnarray}
&&\gamma^{(k)}=\alpha+\beta+\sum_{l=1}^kr_l\,,~~~
\tilde x_m^{(k)}=\frac{\pi}{4}\biggl(\gamma^{(k)}-2\alpha-2\sum_{l=1}^m
r_l\biggr)\,.
\end{eqnarray}
The inner product of $\Phi_{V,M}$ in (\ref{eq:Phi_V^M}) with 
the marginal solution $\psi_{{\rm m},n}$ can be computed in the same way 
as the one done in (\ref{eq:naiseki}) to yield 
\begin{eqnarray}
&&\langle \Phi_{V,M},\psi_{{\rm m},1}\rangle=
\langle {\cal I}|\Phi_{V,M}*\psi_{{\rm m},1}\rangle
=\langle {\cal I}|\psi_{{\rm m},1}*\Phi_{V,M}\rangle
=C_J^{(1)}C_{bc}^{(1)}\,,\\
&&\langle \Phi_{V,M},\psi_{{\rm m},k+1}\rangle=
\langle {\cal I}|\Phi_{V,M}*\psi_{{\rm m},k+1}\rangle
=\langle {\cal I}|\psi_{{\rm m},k+1}*\Phi_{V,M}\rangle\nonumber\\
&&~~~~~~~~~~~~~~~~~~~
=\int_0^{\alpha+\beta}\!\!dr_1\!
\cdots \!\int_0^{\alpha+\beta}\!\!dr_k
\left(\frac{-\pi}{\gamma^{(k)}}\right)^kC_{J}^{(k+1)}C_{bc}^{(k+1)}\,,
\label{eq:phi_cMJ}
\end{eqnarray}
where
\begin{eqnarray}
&&C_J^{(1)}=\sum_{m,n}\zeta_{mn}\,
{}_{\rm mat}\langle 0|\tilde V_m\biggl(\frac{2iM}{\alpha+\beta
}\pm \frac{\pi}{2}\biggr)
\tilde V_n\biggl(\frac{-2iM}{\alpha+\beta}\pm \frac{\pi}{2}\biggr)
\tilde J\biggl(
\frac{\pi(\beta-\alpha)}{2(\alpha+\beta)}
\biggr)|0\rangle_{\rm mat}\,,\\
&&C_J^{(k+1)}=\sum_{m,n}\zeta_{mn}\,
{}_{\rm mat}\langle 0|\tilde V_m\biggl(\frac{2iM}{\gamma^{(k)}
}\pm \frac{\pi}{2}\biggr)
\tilde V_n\biggl(\frac{-2iM}{\gamma^{(k)}}\pm \frac{\pi}{2}
\biggr)\prod_{m=0}^k
\tilde J\biggl(\frac{2\tilde x_m^{(k)}}{\gamma^{(k)}}
\biggr)|0\rangle_{\rm mat}\,,
\label{eq:C_J}
\end{eqnarray}
coming form the matter sector and
\begin{eqnarray}
&&C_{bc}^{(1)}=\frac{i}{2}\sinh\frac{4M}{\alpha+\beta}
\left(\cosh\frac{4M}{\alpha+\beta}-\cos\frac{2\pi \alpha}{\alpha+\beta}
\right),
\\
&&C_{bc}^{(k+1)}=\frac{i}{2\gamma^{(k)}}\biggl(
(\alpha+\beta)\sinh\frac{4M}{\gamma^{(k)}}\cosh\frac{4M}{\gamma^{(k)}}
-\sinh\frac{4M}{\gamma^{(k)}}\biggl(
\alpha\cos\frac{2\pi\beta}{\gamma^{(k)}}+
\beta\cos\frac{2\pi\alpha}{\gamma^{(k)}}\biggr)\nonumber\\
&&~~~~~~~~~~~
-\frac{4M}{\pi}\cosh\frac{4M}{\gamma^{(k)}
}\sin\frac{\pi(\alpha+\beta)}{\gamma^{(k)}}
\cos\frac{\pi(\alpha-\beta)}{\gamma^{(k)}}
+\frac{2M}{\pi}\sin\frac{2\pi(\alpha+\beta)}{\gamma^{(k)}}
\biggr)
\end{eqnarray}
from the ghost sector.

Let us now be more specific by choosing the nonsingular 
current $J=i\partial X^+$, which is the light-cone direction, 
for the marginal solution 
$\Psi^{(\alpha,\beta)}(\lambda_{\rm m}\hat\psi_{\rm m})$
and consider the closed string state 
$\zeta_{mn}V_m(i)V_n(-i)=\zeta_{mn}\partial{X}^{m}(i)\partial{X}^{n}(-i)$. 
Note that there are no constraints on the polarization $\zeta_{mn}$, 
since it carries no momentum due to the fact that the marginal solution 
also carries no momentum. 
By a nonsingular current, we mean that the OPE $J(y)J(z)$ has no 
singularities as $y\to{z}$. 
Since, in order for the specific solution with $J=i\partial X^+$ 
to give the nonvanishing $C_J^{(k)}$, one needs the same number of $X^{-}$ 
as the one of $J$, one can see that only $C_J^{(2)}$ can be nonzero for 
the closed string state with $\zeta_{--}\not=0$. 
As for $C_J^{(2)}$, since $V_m \sim \partial X^m$ and $J=i\partial X^+$ are 
both primary fields of conformal dimension one, their OPE gives 
\begin{eqnarray}
 V_m(y)J(z)&\sim&\frac{v_{m,J}}{(y-z)^2}+{\rm finite},
 ~~~{\rm for}\ y\to z.
\end{eqnarray}
Therefore, it follows from (\ref{eq:C_J}) for $k=1$ that 
\begin{eqnarray}
 C_J^{(2)}&=&{}_{\rm mat}\langle 0|0\rangle_{\rm
  mat}\sum_{m,n}\zeta_{mn}v_{m,J}v_{n,J}\nonumber\\
&&\times
\frac{
(\cos\frac{\pi(\alpha-\beta)}{\gamma^{(1)}}
-\cos\frac{\pi(\alpha+\beta)}{\gamma^{(1)}}\cosh\frac{4M}{\gamma^{(1)}})^2
-(\sin\frac{\pi(\alpha+\beta)}{\gamma^{(1)}}\sinh\frac{4M}{\gamma^{(1)}})^2}
{\left((\cos\frac{\pi(\alpha-\beta)}{\gamma^{(1)}}
-\cos\frac{\pi(\alpha+\beta)}{\gamma^{(1)}}\cosh\frac{4M}{\gamma^{(1)}})^2
+(\sin\frac{\pi(\alpha+\beta)}{\gamma^{(1)}}
\sinh\frac{4M}{\gamma^{(1)}})^2\right)^2}, 
\end{eqnarray}
and thus for large $M$ 
\begin{eqnarray}
 C_J^{(2)}&\sim&\biggl(8{}_{\rm mat}\langle 0|0\rangle_{\rm
  mat}\sum_{m,n}\zeta_{mn}v_{m,J}v_{n,J}\biggr)
e^{-\frac{8M}{\gamma^{(1)}}}\cos\frac{2\pi(\alpha+\beta)}{\gamma^{(1)}}.
\label{eq:CJ2}
\end{eqnarray}
On the other hand, the corresponding ghost contribution
in $\langle\Phi_{\zeta},\psi_{{\rm m},2}\rangle$
cancels the above exponential factor in (\ref{eq:CJ2}), because, 
for large $M$, 
\begin{eqnarray}
 C_{bc}^{(2)}&\sim&
\frac{i(\alpha+\beta)}{8\gamma^{(1)}}e^{\frac{8M}{\gamma^{(1)}}}\,.
\end{eqnarray}
Combining them, one obtains 
\begin{eqnarray}
 \langle\Phi_{\zeta},\psi_{{\rm
  m},2}\rangle&=&\lim_{M\to
  \infty}\langle\Phi_{\zeta,M},\psi_{{\rm
  m},2}\rangle\nonumber\\
&=&\frac{-i\pi}{4}(\alpha+\beta)
{}_{\rm mat}\langle 0|0\rangle_{\rm
  mat}\!\sum_{m,n}\zeta_{mn}v_{m,J}v_{n,J}\!
\int_0^{\alpha+\beta}\!\!dr\!
\left(\frac{2}{\alpha+\beta+r}\right)^2
\cos\frac{2\pi(\alpha+\beta)}{\alpha+\beta+r}\nonumber\\
&=&-i\pi{}_{\rm mat}\langle 0|0\rangle_{\rm
  mat}\!\sum_{m,n}\zeta_{mn}v_{m,J}v_{n,J}\!
\int_0^1dx\frac{1}{(1+x)^2}\cos\frac{2\pi}{1+x}.
\end{eqnarray}
By performing the integration over $x$, one can see that it is zero, 
{\it i.e.}, $\langle\Phi_{\zeta},\psi_{{\rm{m}},2}\rangle=0$. 
Therefore, one finds that the gauge invariant overlap 
for the marginal solution 
\begin{eqnarray}
 {\cal O}_{\zeta}\left(\Psi^{(\alpha,\beta)}(\lambda_{\rm
m}\hat\psi_{\rm m})\right)&=&\sum_{n=1}^{\infty}\lambda_{\rm m}^n
\langle\Phi_{\zeta_{\mu\nu}},\psi_{{\rm
  m},n}\rangle=\lambda_{\rm m}^2\langle\Phi_{\zeta_{\mu\nu}},\psi_{{\rm
  m},2}\rangle=0\,.
\label{eq:O_zeta_marginal}
\end{eqnarray}

Incidentally, the action for the marginal solution (\ref{eq:marginal_ab}) 
hasn't explicitly been computed yet. However, in order to interpret 
the solution physically, it is worth computing. 
In fact, substituting (\ref{eq:marginal_ab}) into the action, one obtains 
\begin{eqnarray}
&&S[\Psi^{(\alpha,\beta)}(\lambda_{\rm m}\hat\psi_{\rm m})]=
\frac{1}{6g^2}\sum_{k,lm,\ge 0}\lambda_{\rm
m}^{k+l+m+3}S_{klm}^{(\alpha,\beta)},
\end{eqnarray}
where 
\begin{eqnarray}
&&S_{klm}^{(\alpha,\beta)}=(-1)^{k+l+m}
\langle {\cal I}|
(\hat{\psi}_{\rm m}A^{(\alpha+\beta)})^k\hat{\psi}_{\rm m}
P_{\alpha+\beta}
(\hat{\psi}_{\rm m}A^{(\alpha+\beta)})^l\hat{\psi}_{\rm m}
P_{\alpha+\beta}(\hat{\psi}_{\rm m}A^{(\alpha+\beta)})^m\hat{\psi}_{\rm m}
P_{\alpha+\beta}\rangle\,.\nonumber\\
\end{eqnarray}
Recalling that $A^{(\gamma)}=\frac{\pi}{2}\int_0^{\gamma}d\alpha
B_1^L\hat{U}_{\alpha+1}|0\rangle$, 
one finds that 
\begin{eqnarray}
S_{klm}^{(\alpha,\beta)}&=&
\int_0^{\alpha+\beta}\!\!dr_1\!\cdots\!
\int_0^{\alpha+\beta}\!\!dr_{k+l+m}
\left(\frac{-\pi}{\gamma^{(klm)}}\right)^{k+l+m}\!
C_J^{(klm)}C_{bc}^{(klm)},
\end{eqnarray}
where 
\begin{eqnarray}
C_J^{(klm)}&=&{}_{\rm mat}\langle 0|\prod_{j_1=0}^k{\tilde
 J}(\tilde x_{j_1}^{(klm)})
\prod_{j_2=0}^l{\tilde J}(\tilde y_{j_2}^{(klm)})
\prod_{j_3=0}^m{\tilde J}(\tilde z_{j_3}^{(klm)})|0\rangle_{\rm mat}\,,
\label{eq:C_Jklm_JJJ}
\\
C_{bc}^{(klm)}&=&\frac{\alpha+\beta
}{\gamma^{(klm)}}\left(
\sin\frac{\pi (\alpha+\beta)
}{\gamma^{(klm)}}\right)^2\biggl(
\sin\frac{2\pi (\alpha+\beta+\sum_{j=1}^{k}r_j)
}{\gamma^{(klm)}}\nonumber\\
&&+
\sin\frac{2\pi (\alpha+\beta+\sum_{j=k+1}^{k+l}r_j)
}{\gamma^{(klm)}}
+
\sin\frac{2\pi (\alpha+\beta+\sum_{j=k+l+1}^{k+l+m}r_j)
}{\gamma^{(klm)}}\biggr),
\end{eqnarray}
along with 
\begin{eqnarray}
&&\gamma^{(klm)}= 3(\alpha+\beta)+\sum_{j=1}^{k+l+m}r_j\,,
~~~\tilde x_j^{(klm)}
=\frac{\pi}{4}(2(\alpha+\beta)-\sum_{i=1}^{j}r_i
+\sum_{i=j+1}^{k+l+m}r_i)\,,\\
&&\tilde y_j^{(klm)}
=\frac{\pi}{4}(-\sum_{i=1}^{j+k}r_i+\sum_{i=j+k+1}^{k+l+m}r_i)\,,
~~~\tilde z_j^{(klm)}=\frac{\pi}{4}(-2(\alpha+\beta)
-\sum_{i=1}^{j+k+l}r_i+\sum_{i=j+k+l+1}^{k+l+m}r_i)\,.\nonumber
\end{eqnarray}
Since, for the matter current $J=i\partial{X}^{+}$, 
there is no $X^{-}$ inserted in the correlator (\ref{eq:C_Jklm_JJJ}), 
$C_J^{(klm)}$ must be zero. One can thus find that 
\begin{eqnarray}
\label{eq:S_marginal_0}
&&S[\Psi^{(\alpha,\beta)}(\lambda_{\rm m}\hat\psi_{\rm m})]=0\,.
\end{eqnarray}

It is known \cite{Kishimoto:2007bb} that 
the marginal solution (\ref{eq:marginal_ab}) with $J=i\partial X^+$ 
can be put in the form of pure gauge, na\"ively. 
In fact, since both of the gauge invariant overlap 
(\ref{eq:O_zeta_marginal}) and the action (\ref{eq:S_marginal_0}) 
for the solution (\ref{eq:marginal_ab}) are vanishing, 
our results are consistent with it.

\section{
Discussions
\label{sec:Dis}}

In this paper, the gauge invariant overlaps for the three solutions;
Schnabl's solution \cite{Schnabl_tach}, the level truncated 
solution in the Siegel gauge 
\cite{Sen:1999nx,Moeller:2000xv,Gaiotto:2002wy}, 
and the marginal solution 
\cite{Schnabl:2007az, Kiermaier:2007ba, Kishimoto:2007bb}, 
have been computed to give the expected results; 
non-zero values for the first two solutions and zero 
for the last one.

The vacuum energy for Schnabl's solution $\Psi_{\lambda}$ with $\lambda=1$ 
gives the correct D-brane tension and provides an important evidence for 
Sen's conjecture on tachyon condensation. 
Although $\Psi_{\lambda}$ with any $\lambda$ is a classical solution to the 
equation of motion, all the solutions except for the one with $\lambda=1$ 
can be gauged away to be trivial. 
Therefore, in addition to the vacuum energy, 
the results in this paper give another evidence that 
the Schnabl's solution $\Psi_{\lambda}$ is nontrivial
only for $\lambda=1$. In particular, it has been confirmed 
analytically and numerically.

Furthermore, our results for the level truncated solution in the 
Siegel gauge \cite{Sen:1999nx,Moeller:2000xv,Gaiotto:2002wy} give 
another evidence that it may be a gauge equivalent to Schnabl's solution.
The gauge invariant overlaps give other gauge invariant observables 
than the action itself and can distinguish gauge inequivalent solutions. 
Therefore, the results in this paper yield an interesting match between 
the two solutions.

Although the gauge invariant overlaps for the three solutions were 
computed in this paper, the physical meaning of them is quite obscure. 
The gauge invariant overlaps were originally introduced to give the coupling 
of an on-shell closed string state with open string fields in open string 
field theory \cite{Zwiebach:1992bw}. Therefore, the closed string state 
plays a role of the source term for a dynamical open string field.
The gauge invariant overlaps thus seem the couplings of the 
on-shell closed string states with the classical solutions, 
or in other words a kind of the back reaction of the classical 
solutions to the closed string sector. 
Anyhow, it would be interesting to make it clearer.

\bigskip
{\sl Note added:} after writing up the manuscript, we are aware of 
a paper \cite{Ellwood:2008jh} appearing on the arXiv, 
whose results have substantial overlap with ours.

\section*{Acknowledgments}

We are indebted to Taichiro Kugo for collaboration at the early stages of 
this work and sharing his insight with us on the work.
I.~K. would like to thank Seiji Terashima for valuable discussions. 
The work of T.~K. was supported in part by a Grant-in-Aid (\#19540268) 
from the MEXT of Japan.
The work of I.~K. was supported in part by the Special Postdoctoral 
Researchers Program at RIKEN and Grant-in-Aid for Young Scientists (\#19740155) 
from the MEXT of Japan.
The work of T.~T. was supported in part by a Grant-in-Aid for 
Young Scientists (\#18740152) from the MEXT of Japan.
The level truncation calculations based on {\sl Mathematica} were 
carried out partly on the computer {\it sushiki} at Yukawa Institute 
for Theoretical Physics in Kyoto University.

\appendix

\section{Oscillator Expressions for the On-shell Closed Tachyon and 
Dilaton States
\label{sec:Ginv_closed}}

A simple example 
of on-shell closed string states in the open string Hilbert space
is the tachyon state and its oscillator expression is given 
\cite{Takahashi:2003kq} by 
\begin{eqnarray}
\label{eq:closed_tach}
\Phi_{k}&=&\frac{1}{4}\,e^{E_{\rm m}+E_{\rm
 gh}}c_0c_1|0\rangle\,,
\\
E_{\rm m}&=&-\sum_{n=1}^{\infty}
\frac{(-1)^n}{2n}\alpha_{-n}\cdot\alpha_{-n}
-\sum_{n=1}^{\infty}\frac{2i\sqrt{2\alpha'}(-1)^n
}{2n-1}k_i\alpha_{-2n+1}^i\,,
\label{eq:E_rmm}\\
E_{\rm gh}&=&\sum_{n=1}^{\infty}(-1)^nc_{-n}b_{-n}\,,
\label{eq:E_rmgh}
\end{eqnarray}
where we set the momentum along the  Neumann direction ($\mu=0,1,\cdots
,p$) to be zero.
The string coordinates along the Dirichlet direction $X^i(z,\bar z)$
($i=p+1,\dots, 25$) are given by $X^i(z,\bar z)=(X(z)-X(\bar{z}))^i/2$
with
\begin{eqnarray}
&&X^M(z)=x^M-i\sqrt{2\alpha'}\alpha_0^M\log z+
i\sqrt{2\alpha'}\sum_{n\ne 0}{1\over n}\alpha_{n}^Mz^{-n},\\
&&[\alpha_n^M,\alpha_m^N]=n\delta_{n+m,0}\eta^{MN},
~~~[x^M,\alpha_0^N]=i\sqrt{2\alpha'}\eta^{MN}\,.
\end{eqnarray}
In fact, (\ref{eq:closed_tach}) is obtained by acting 
\begin{eqnarray}
 V_{k}(i)
=\frac{1}{4i}c(i)c(-i):\!e^{ik\cdot X(i,-i)}\!:
\label{eq:closedstringtachyon}
\end{eqnarray}
on the identity state $|{\cal I}\rangle$ :
\begin{eqnarray}
V_k(i)|{\cal I}\rangle
&\equiv&\frac{1}{4i}
\lim_{\theta\to \frac{\pi}{2}}
c(e^{i\theta})c(e^{-i\theta}):\!
e^{ik\cdot X(e^{i\theta},e^{-i\theta})}\!\!:\!|{\cal
I}\rangle\nonumber\\
&=&i\lim_{\theta\to \frac{\pi}{2}}
c(e^{i\theta})\!:\!e^{\frac{i}{2}k\cdot X(e^{i\theta})}
\!\!:c(e^{-i\theta})\!
:\!e^{-{\frac{i}{2}}k\cdot X(e^{-i\theta})}
\!\!:\!|{\cal I}\rangle
=\Phi_k,
\label{eq:V_k2Phi_k}
\end{eqnarray}
where we have used the formulae 
\begin{eqnarray}
&&|{\cal I}\rangle=e^{E_{{\cal I},{\rm m}}+E_{{\cal I},{\rm
 gh}}}|0\rangle\,,
\label{eq:identity_osc}
\\
&&E_{{\cal I},{\rm m}}= -\frac{1}{2}\sum_{n=1}^\infty\frac{(-1)^n }{n}
\alpha_{-n} \cdot \alpha_{-n}\,,
\label{eq:E_Irmm}
\\
&&E_{{\cal I},{\rm gh}}=
\sum_{n=2}^{\infty}(-1)^n c_{-n}b_{-n}
-2 c_0\sum_{n=1}^{\infty}(-1)^n b_{-2n}
-(c_1-c_{-1})\sum_{n=1}^{\infty}(-1)^nb_{-(2n+1)}\,,\\
&&\sum_{n=-\infty}^{\infty}c_ne^{-in\theta}|{\cal I}\rangle=
\biggl[ic_0\tan\theta+c_1\frac{1}{2\cos\theta}+c_{-1}\frac{1+2\cos
2\theta}{2\cos\theta}\nonumber\\
&&~~~~~~~~~~~~~~~~~~~~~~~~~
+2\sum_{n=1}^{\infty}(c_{-2n} i\sin 2n\theta+c_{-2n-1}\cos(2n+1)\theta)
\biggr]|{\cal I}\rangle\,,
\end{eqnarray}
and the on-shell condition $\alpha'k^2=4$.
By a straightforward calculation, one finds that 
\begin{eqnarray}
 Q_{\rm B}|\Phi_k\rangle
&=&4(\alpha'k^2-4)\sum_{m=1}^{\infty}(-1)^m m c_{-2m}|\Phi_k\rangle,
\end{eqnarray}
which vanishes for  $\alpha'k^2=4$. The on-shell condition guarantees
that the conformal dimension of
$:\!e^{\pm {\frac{i}{2}}k\cdot X(\pm i)} \!\!:$ is one.

In (\ref{eq:V_k2Phi_k}), we have introduced  $\theta$ to make
the computation well-defined
because there are subtleties concerned with the divergence 
at the point $z=i$ (or $\theta=\pi/2$). The
regularization parameter $\theta$ corresponds to $M$ in the sliver frame
introduced in (\ref{eq:Phi_V^M}). They are related as
$\tan(\theta/2)=\tanh M$ which follows from (\ref{eq:hatU_1-1}).
If we use the CFT expression (\ref{eq:Zwiebach_inv}),
we can avoid the subtleties, which are related to the conformal
factor.

The massless closed string state with zero momentum takes
\begin{eqnarray}
 V_{\zeta}(i)&=&\frac{i}{26\cdot (-2\alpha')}
\sum_{M,N}\zeta_{MN}c(i)\partial X^M(i)c(-i)\partial X^N(-i)\,.
\end{eqnarray}
In the same way as (\ref{eq:V_k2Phi_k}), 
one can compute the oscillator expression for $V_{\zeta}(i)|{\cal I}\rangle$ 
as 
\begin{eqnarray}
\label{eq:closed_massless1}
V_{\zeta}(i)|{\cal I}\rangle&=&\frac{1}{52\alpha' i}
\sum_{M,N}\zeta_{MN}
\lim_{\theta\to \frac{\pi}{2}}
  c(e^{i\theta})\partial X^M(e^{i\theta})c(e^{-i\theta})
\partial X^N(e^{-i\theta})|{\cal I}\rangle\equiv \Phi_{\zeta}.
\end{eqnarray}
Using the formula 
\begin{eqnarray}
\partial X^M(e^{i\theta})\partial X^N(e^{-i\theta})|{\cal
 I}\rangle
&=&
 -2\alpha'\biggl[\,\sum_{n,m=1}^{\infty}\alpha_{-n}^N\alpha_{-m}^M
(e^{-in\theta}-(-1)^ne^{in\theta})
(e^{im\theta}-(-1)^ne^{-im\theta})\nonumber\\
&&~~~~~~~
+\eta^{MN}\sum_{n=1}^{\infty}n(e^{-2in\theta}-(-1)^n)
\biggr]|{\cal I}\rangle,
\end{eqnarray}
and defining the summation in the last line as
\begin{eqnarray}
&&\sum_{n=1}^{\infty}n(e^{-2in\theta}-(-1)^n)
=f(\theta)-f(\pi/2)\,,~~~~~
f(\theta)\equiv \sum_{n=1}^{\infty}n e^{-2in\theta}\,,
\end{eqnarray}
with the regularization
\begin{eqnarray}
\label{eq:ftheta_reg}
 f(\theta)&= &\lim_{\epsilon\to +0}
\sum_{n=1}^{\infty}n\,e^{-2in\theta-n\epsilon}=-\frac{1}{4(\sin\theta)^2},
\end{eqnarray}
one obtains 
\begin{eqnarray}
\Phi_{\zeta}&=&\frac{1}{26}\zeta_{MN}
\left(\frac{1}{4}\eta^{MN}
-4\sum_{n,m=1}^{\infty}i^{m-n}mn\alpha_{-m}^M\alpha_{-n}^N
\right)e^{E_{{\cal I},{\rm m}}+E_{\rm gh}}c_0c_1|0\rangle,
\label{eq:closed_massless}
\end{eqnarray}
where 
\begin{eqnarray}
 \lim_{\theta\to \frac{\pi}{2}}\frac{f(\theta)-f(\pi/2)}{(\cos\theta)^2}
&=&-\frac{1}{4}\,,
\label{eq:ftheta-1/4}
\end{eqnarray}
was used. 

In above formula, $E_{{\cal I},{\rm m}}$ and $E_{\rm gh}$ are given by
(\ref{eq:E_Irmm}) and (\ref{eq:E_rmgh}), respectively.
The result of the calculations 
\begin{eqnarray}
&&Q_{\rm B}\left(
\sum_{n,m=1}^{\infty}i^{m-n}mn\alpha_{-m}^M\alpha_{-n}^N
e^{E_{{\cal I},{\rm m}}+E_{\rm gh}}c_0c_1|0\rangle\right)
=\eta^{MN}\sum_{k=1}^{\infty}(-1)^{k+1}kc_{-2k}
e^{E_{{\cal I},{\rm m}}+E_{\rm gh}}c_0c_1|0\rangle\,,\nonumber\\
\\
&&Q_{\rm B}\,e^{E_{{\cal I},{\rm m}}+E_{\rm gh}}c_0c_1|0\rangle
=-16\sum_{k=1}^{\infty}(-1)^kkc_{-2k}
e^{E_{{\cal I},{\rm m}}+E_{\rm gh}}c_0c_1|0\rangle\,,
\end{eqnarray}
makes sure its BRST invariance $Q_{\rm B}|\Phi_{\zeta}\rangle=0$,
which justifies our prescription in (\ref{eq:ftheta_reg}).
In particular, one obtains the dilaton state with zero momentum
\begin{eqnarray}
\Phi_{\eta}&=&\frac{1}{52\alpha'i}
c(i)c(-i)\partial X(i)\cdot\partial X(-i)|{\cal I}\rangle\nonumber\\
&=&
\left(\frac{1}{4}
-\frac{2}{13}\sum_{n,m=1}^{\infty}mn\cos\frac{(m-n)\pi}{2}
\alpha_{-m}\cdot \alpha_{-n}
\right)e^{E_{{\cal I},{\rm m}}+E_{\rm gh}}c_0c_1|0\rangle\,
\label{eq:dilaton}
\end{eqnarray}
by taking $\zeta_{MN}=\eta_{MN}$.

\section{The Shapiro-Thorn Vertex
and Closed String States
\label{sec:ST-vertex}}

We begin with a brief review on the Shapiro-Thorn  $\hat\gamma$ vertex in
\cite{Shapiro:1987ac},
which is defined 
with the conformal maps $h_1(w)=-i(w-1)/(w+1)$
and $h_2(w)=(w-1/w)/2$ 
by the CFT correlator on the upper half plane as 
$\langle\hat\gamma(1_{\rm c},2)|\phi_{\rm c}\rangle_{1_{\rm
c}}|\psi\rangle_2=\langle h_1[\phi_{\rm c}(0,0)]h_2[\psi(0)]\rangle$
by using the LPP method \cite{LPP}, as in (\ref{eq:S-T_LPPdef}).
The holomorphic function $h_1(w)$ for a closed string $1_{\rm c}$
is a map from the unit disk to the upper half plane, 
which satisfies $h_1(0)=i$. The function 
$h_2(w)$ for an open string $2$ is a map from the half unit disk to the
upper half plane, which satisfies $h_2(w)=I\circ f_{\cal I}(w)$.
$I(z)=-1/z$ is the inversion map and the function $f_{\cal I}(w)$ defines 
the identity state. 
If one regards the holomorphic and 
anti-holomorphic part in the closed string sector
$1_{\rm c}$ as two open string sectors, $1,1^*$,
one can construct $\langle\hat\gamma(1_{\rm c},2)|$ 
as a vertex for three open strings.
The conformal map for $1^*$ is then given by $h_{1^*}(w)=i(w-1)/(w+1)$.

By taking the doubling trick and inserting the BRST charge as the contour 
integral of the BRST current $\oint j_{\rm B}(z)$ in the correlator,
one can see the BRST invariance (\ref{eq:BRS_hatgamma_R}) of 
$\langle\hat\gamma(1_{\rm c},2)|$ as  
\begin{eqnarray}
&&\langle\hat\gamma(1_{\rm c},2)|(Q_{\rm B}^{(1)}+\bar Q_{\rm B}^{(1)}
+Q_{\rm B}^{(2)})|\phi_{\rm c}\rangle_{1_{\rm c}}|\psi\rangle_{2}
=\left(\oint_i +\oint_{-i}+\oint_{\infty}\right)\frac{dz}{2\pi i}
\langle j_{\rm B}(z)h_1[\phi_{\rm c}(0,0)]h_2[\psi(0)]\rangle
\nonumber\\
&&=-\oint \frac{dz}{2\pi i}
\langle j_{\rm B}(z)h_1[\phi_{\rm
c}(0,0)]h_2[\psi(0)]\rangle=0\,
\end{eqnarray}
for any $\phi_{\rm c},\psi$.

One can use  $\langle\hat\gamma(1_{\rm c},2)|$ to define 
a gauge invariant overlap ${\cal O}_V(\Psi)$
by $\langle\hat\gamma(1_{\rm c},2)|V_{\rm c}\rangle_{1_{\rm
c}}|\Psi\rangle_2$ as in (\ref{eq:O_Vgamma})
with  $V_{\rm c}(z,\bar z)$, which is BRST invariant 
and a primary field  of conformal dimension $(0,0)$ and the ghost number two.

The relation (\ref{eq:com_on_closed}) can be shown as follows.
With a formal expression 
$|R(3,2)\rangle=\sum_r|\phi_r\rangle_3|\phi^c_r\rangle_2$ 
of the reflector with a complete set of the Hilbert space 
with $\langle \phi^c_r,\phi_s\rangle=\delta_{r,s}$,
${\cal O}_V(\psi*\varphi)$ can be computed in the same way 
as in \cite{KO}\footnote{
Here, $\langle V_3(3,4,5)|$ is the 3-string vertex which is defined by
the three maps $f^{(3)}_{1,2,3}(z)$. $F_1$ and $\hat{F}_2$
given in \cite{KO}  are 
the re-smoothing maps for the gluing procedure. ${\cal R}_{\pi/2}^{-1}$
is the inverse of
${\cal R}_{\pi/2}(z)=F_1\circ f_3^{(3)}(z)=(z+1)/(-z+1)$.
} to yield 
\begin{eqnarray}
{\cal O}_V(\psi*\varphi)
&=&\langle \hat{\gamma}(1_c,2)|V_{\rm c}\rangle_{1_c}
\langle
V_3(3,4,5)|R(3,2)\rangle|\psi\rangle_4|\varphi\rangle_5\nonumber\\
&=&\sum_{r}\langle
V_3(3,4,5)|\phi_r\rangle_3|\psi\rangle_4|\varphi\rangle_5
\langle \hat{\gamma}(1_c,2)|V_{\rm c}\rangle_{1_c}|
\phi_r^c\rangle_2\nonumber\\
&=&\sum_{r}\langle
 f_1^{(3)}[\varphi(0)]\,f_2^{(3)}[\phi_r(0)]\,
f_3^{(3)}[\psi(0)]\rangle\,\langle I\circ h_1[V_{\rm c}(0,0)]\,f_{\cal
I}[\phi_r^c(0)]\rangle\nonumber\\
&=&
\left\langle
F_1\circ f_3^{(3)}[\psi(0)]\,
F_1\circ f_1^{(3)}[\varphi(0)]\,
\hat F_2\circ I\circ h_1[V_{\rm c}(0,0)]
\right\rangle\nonumber\\
&=&\left\langle\psi(0)\,I[\varphi(0)]\,
{\cal R}_{\pi/2}^{-1}\circ \hat F_2\circ I\circ h_1[V_{\rm c}(0,0)]
\right\rangle\,.
\end{eqnarray}
Using ${\cal R}_{\pi/2}^{-1}\circ \hat F_2\circ I\circ h_1(0)=i$ and 
the condition that $V_{\rm c}(z,\bar z)$ is a primary field 
of conformal dimension $(0,0)$, one obtains  
\begin{eqnarray}
{\cal O}_V(\psi*\varphi)
=\left\langle\psi(0)\,I[\varphi(0)]\,V_{\rm c}(i,-i)\right\rangle
=\left\langle\varphi(0)\,I[\psi(0)]\,V_{\rm c}(i,-i)\right\rangle
={\cal O}_V(\varphi*\psi)\,.
\label{eq:comm_LPP}
\end{eqnarray}
Note that $\partial ({\cal R}_{\pi/2}^{-1}\circ \hat F_2\circ
I\circ h_1)(0)=\infty$, which implies that ${\cal O}_V(\psi*\varphi)$
is not well-defined, if  $V_{\rm c}(z,\bar z)$ has nonzero conformal dimension, 
because of its conformal factor.

It may be useful to give the explicit form of the vertex 
$\langle\hat\gamma(1_{\rm c},2)|$ 
in terms of the oscillators.
{}From the three maps $h_1,h_{1^*},h_2$, one can compute the Neumann
coefficients $\bar N^{rs}_{nm}$ as
\begin{eqnarray}
&&\bar{N}^{rs}_{nm}=\frac{1}{nm}\oint_0{dw_r\over 2\pi i}
\oint_0{dw_s\over 2\pi i}\frac{w_r^{-n}w_s^{-m}
h'_r(w_r)h'_s(w_s)}{
(h_r(w_r)-h_s(w_s))^2},~~~(n,m\ge 1),
\\
&&\bar{N}^{rs}_{n0}=\bar{N}^{sr}_{0n}=\frac{1}{n}
\oint_0{dw_r\over 2\pi i}\frac{w_r^{-n}h'_r(w_r)}{h_r(w_r)-h_s(0)},
~~~(n\ge 1),\\
&&\bar{N}^{rs}_{00}=\log|h_r(0)-h_s(0)|,~~(r\ne s),~~~~~
\bar{N}^{rr}_{00}=\log|h'_r(0)|,
\end{eqnarray}
for the matter sector and also $N^{(g)rs}_{~~~nm}, M^r_{im}$ as 
\begin{eqnarray}
&& N^{(g)rs}_{~~~nm}=\oint_0{dw_r\over 2\pi i}
\oint_0{dw_s\over 2\pi i}
(h_r'(w_r))^2(h_s'(w_s))^{-1}
{-w_r^{-n+1}w_s^{-m-2}\over h_r(w_r)-h_s(w_s)}\,,
~~~~(n\ge 2,~m\ge -1),\nonumber
\\
&&M^r_{in}=\oint_0{dw_r\over 2\pi i}(h'_r(w_r))^{-1}w_r^{-n-2}
(h_r(w_r))^{i+1},~~~~~(i=-1,0,1,~~~n\ge -1).
\end{eqnarray}
for the ghost sector \cite{LPP}. 
Using the relation
\begin{eqnarray}
&&{}_{1,1^*,2}\langle 3|\prod_{i={-1}}^1
\left(\sum_r\sum_{m\ge -1}M^r_{im}b_m^{(r)}\right)\nonumber\\
&&=\left(\det_{r,i}M^r_{i,-1}\right)
{}_{1,1^*,2}\langle \tilde 1|\exp\left(\sum_{r,s,i}\sum_{m\ge 0}
c^{(r)}_1((M_{,-1})^{-1})_{ri}M^{s}_{im}b^{(s)}_m
\right)
\end{eqnarray}
in the ghost sector,
where $\langle \tilde{1}|c_1\equiv \langle 3|\equiv \langle
0|c_{-1}c_0c_1$,
one obtains the explicit form of 
the vertex  $\langle\hat\gamma(1_{\rm c},2)|$ as 
\begin{eqnarray}
 \langle \hat{\gamma}(1_c,2)|&=&{1\over 4}\int{d^{26}p_1\over (2\pi)^{26}}
\int{d^{26}\bar p_1\over (2\pi)^{26}}
\int{d^{26}p_2\over (2\pi)^{26}}(2\pi)^{26}\delta^{26}(p_1+\bar p_{1}+p_2)
\nonumber\\
&&~~\times{}_{1_c}\langle p_1;\bar p_1|c_{-1}c_0\bar c_{-1}\bar c_0\,\,
{}_2\langle p_2|c_{-1}c_0\,e^{E(1_c,2)},
\label{eq:hatgamma}
\end{eqnarray}
where
\begin{eqnarray}
&&E(1_c,2)=
\alpha_0^{(2)}\cdot\alpha_0^{(2)}\log 2
- \sum_{n=1}^{\infty}\frac{2}{n}((-i)^n\alpha_0^{(1)}+i^n\bar
\alpha_0^{(1)})\!\cdot \!\alpha_n^{(2)}
-\alpha_0^{(2)}\!\cdot\!\sum_{n=1}^{\infty}{(-1)^n\over
 n}(\alpha^{(1)}_{n}+\bar{\alpha}^{(1)}_{n})\nonumber\\
&&~~~~-{1\over 2}\sum_{n=1}^{\infty}{(-1)^n\over
 n}\alpha_n^{(2)}\!\cdot\!\alpha_n^{(2)}
-\sum_{n=1}^{\infty}{1\over n}\alpha^{(1)}_n\cdot
\bar{\alpha}^{(1)}_n
-\sum_{n,m=1}^{\infty}
{(-1)^m\over m}\eta^{2m}_n
\alpha_n^{(2)}\!\cdot\!((-i)^n\alpha^{(1)}_m+i^n\bar{\alpha}^{(1)}_m)
\nonumber\\
&&~~~~-\sum_{n=1}^{\infty}\left((-1)^nc^{(2)}_nb^{(2)}_n
+c_n^{(1)}\bar{b}_n^{(1)}+\bar{c}_n^{(1)}b^{(1)}_n+(-1)^n
(c^{(1)}_n-\bar{c}^{(1)}_n)(b^{(1)}_0-\bar{b}^{(1)}_0)\right)
\nonumber\\
&&~~~~-\sum_{m=1}^{\infty}\sum_{n=0}^{\infty}\biggl(
2(-1)^n(\eta^m_{2n+1}-\eta^m_{2n-1})c_m^{(2)}((-i)^mb^{(1)}_n
+i^m\bar{b}_n^{(1)})\\
&&~~~~~~~~~~~~
+{(-1)^m\over 4}(\eta^{2m}_{n+1}-\eta^{2m}_{n-1}+\delta_{n,1})
((-i)^nc^{(1)}_m+i^n\bar{c}^{(1)}_m)b^{(2)}_n
\biggr),
\nonumber
\end{eqnarray}
and the coefficients $\eta^k_n$ are given by the generating function 
\begin{eqnarray}
 &&\left(1+x\over 1-x\right)^k= \sum_{n=0}^{\infty}\eta^k_n x^n\,.
\label{eq:def_eta_coeff}
\end{eqnarray}
Here, as our convention $p_r=\alpha_0^{(r)}/\sqrt{2\alpha'}$, and
we regard $\alpha_0^{(1)},\bar \alpha_0^{(1)}$ as
independent modes with the normalization 
${}_{1_c}
\langle p_1;\bar p_1|p_1';\bar p_1'\rangle_{1_c}
=(2\pi)^{26}\delta^{26}(p_1-p_1')
(2\pi)^{26}\delta^{26}(\bar p_1-\bar p_1')$
for the matter zero modes in the closed string sector.

The mapping of a closed string state $|\Phi_{\rm c}\rangle$
in the closed string Hilbert space
by $ \langle \hat{\gamma}(1_{\rm c},2)|$ (\ref{eq:hatgamma}) 
gives  the state 
$ \langle \hat{\gamma}(1_{\rm c},2)|\Phi_{\rm c}\rangle_{1_{\rm c}}$ 
in the open string Hilbert space. 
In particular, $\langle
 \hat{\gamma}(1_{\rm c},2)|\Phi_{\rm c}\rangle_{1_{\rm c}}$
is BRST invariant if $|\Phi_{\rm c}\rangle$ is so
thanks to $(\ref{eq:BRS_hatgamma_R})$.
In the following, we will explicitly see a few examples of the images 
by the map $|\Phi_{\rm c}\rangle\mapsto \langle
 \hat{\gamma}(1_{\rm c},2)|\Phi_{\rm c}\rangle_{1_{\rm c}}$.

\paragraph{Closed tachyon state}
In the closed string Hilbert space, the closed tachyon state is 
\begin{eqnarray}
\label{eq:tachyon_c}
{}|\Phi_{(p,k)}\rangle&=&
c_1\bar c_1|(p^{\mu}/2,k^i/2);(p^{\mu}/2,-k^i/2)\rangle
\end{eqnarray}
where T-dual transformation is taken for the
directions of $k^i$. Namely, it corresponds to 
$c(z)\bar c(\bar z)\!:\!e^{ip_{\mu}X^{\mu}(z,\bar z)+ik_iX^i(z,\bar z)}\!:$
with $X^{\mu}(z,\bar z)=(X^{\mu}(z)+\bar X^{\mu}(\bar z))/2$
~($\mu=0,1,\cdots,p$) and
$X^i(z,\bar z)=(X^i(z)-\bar X^i(\bar z))/2$~
($i=p+1,\cdots, 25$).
The BRST invariance $(Q_{\rm B}+\bar{Q}_{\rm
B})|\Phi_{(p,k)}\rangle=0$ is satisfied by
the mass shell condition: $p^2+k^2=4/\alpha'$.
By contracting 
with $ \langle \hat{\gamma}(1_{\rm c},2)|$, we get
\begin{eqnarray}
 \langle \hat\gamma(1_{\rm c},2)|\Phi_{(p,k)}\rangle_{1_{\rm c}}&=&
\frac{1}{4}\,{}_2\langle (-p^{\mu},0)|c_{-1}c_0e^{E_{\tilde{\cal
I}}+E(p,k)}\,,\\
E_{\tilde{\cal I}}&=&-\sum_{n=1}^{\infty}
(-1)^n\left(\frac{1}{2n}\alpha_n\!\cdot\alpha_n
+c_{n}b_{n}\right)\,,
\label{eq:E_tildeI}\\
E(p,k)&=&2\alpha'p^2\log
 2-\sum_{n=1}^{\infty}\frac{\sqrt{2\alpha'}(-1)^n}{n}
p_{\mu}\alpha_{2n}^{\mu}
-\sum_{n=1}^{\infty}\frac{2i\sqrt{2\alpha'}(-1)^n
}{2n-1}k_i\alpha_{-2n+1}^i,~~~~~~~~
\label{eq:E_pk}
\end{eqnarray}
which is equal to the tachyon state $\langle T;p,k|$
in open string Hilbert space given in \cite{Takahashi:2003kq}.
By taking $p^{\mu}=0$ and BPZ conjugation, we reproduce
(\ref{eq:closed_tach}):
\begin{eqnarray}
&&\langle
 \hat\gamma(1_{\rm c},2)|\Phi_{(0,k)}\rangle_{1_{\rm c}}|R(2,3)\rangle
=|\Phi_k\rangle_3\,.
\end{eqnarray}

\paragraph{Closed massless state}
In the closed string Hilbert space, closed massless state 
with polarization $\zeta_{MN}$ is
\begin{eqnarray}
{}|\Phi_{\zeta,(p,k)}\rangle&=&-\frac{1}{26}
\alpha_{-1}^M\bar{\alpha}_{-1}^N
c_1\bar{c}_1|(p/2,k/2);(p/2,-k/2)\rangle\zeta_{MN}\,.
\label{eq:Phizeta_pk}
\end{eqnarray}
The condition for the BRST invariance: $(Q_{\rm B}+\bar Q_{\rm
B})|\Phi_{\zeta,(p,k)}\rangle=0$ is 
\begin{eqnarray}
 p^2+k^2=0\,,~~~~~p^{\mu}\zeta_{\mu N}+k^i\zeta_{i N}=0\,,~~~~
\zeta_{M \nu}p^{\nu}-\zeta_{M j}k^j=0\,.
\label{eq:massless_BRST}
\end{eqnarray}
The image of $\langle \hat\gamma(1_{\rm c},2)|$ is computed as
\begin{eqnarray}
\langle \hat\gamma(1_{\rm c},2)|\Phi_{\zeta,(p,k)}\rangle_{1_{\rm c}}&=&
\frac{1}{26}\,{}_2\langle (-p^{\mu},0)|c_{-1}c_0e^{E_{\tilde{\cal
I}}+E(p,k)}\biggl(\frac{1}{4}\eta^{MN}\zeta_{MN}-4E^M \bar E^N\zeta_{MN}
\nonumber\\
&&
+\sqrt{2\alpha'}p^{\mu}\bar E^N\zeta_{\mu N}
+\sqrt{2\alpha'}p^{\nu}E^M
\zeta_{M\nu}-\frac{\alpha'}{2}p^{\mu}p^{\nu}\zeta_{\mu\nu}
\biggr)\,,\\
&&E^M=\sum_{n=1}^{\infty}i^{-n}n\alpha_n^M\,,~~~
\bar E^N=\sum_{n=1}^{\infty}i^{n}n\alpha_n^N\,,
\end{eqnarray}
where $E_{\tilde{\cal I}}$ and $E(p,k)$ are given in (\ref{eq:E_tildeI})
and (\ref{eq:E_pk}). The BPZ conjugation of the above with
$(p^{\mu}=0,k^i=0)$ yields (\ref{eq:closed_massless}):
\begin{eqnarray}
&&\langle \hat{\gamma}(1_{\rm c},2)|
  \Phi_{\zeta,(0,0)}\rangle_{1_c}|R(2,3)\rangle
=|\Phi_{\zeta}\rangle_3\,.
\end{eqnarray}
For nonzero momentum, the conditions in (\ref{eq:massless_BRST})
cannot be satisfied for $\zeta_{MN}=\eta_{MN}$.
Instead of (\ref{eq:Phizeta_pk}), we define 
\begin{eqnarray}
{}|\Phi_{D,(p,k)}\rangle&\equiv&
(\alpha_{-1}\!\cdot\!\bar{\alpha}_{-1}
c_1\bar{c}_1
-(c_{-1}c_1-\bar{c}_{-1}\bar{c}_1))|(p/2,k/2);(p/2,-k/2)\rangle
\end{eqnarray}
and then it becomes BRST invariant by the massless condition: $p^2+k^2=0$.
However, its image with massless momentum in the open string Hilbert space:
$\langle \hat{\gamma}(1_{\rm c},2)|
  \Phi_{D,(p,k)}\rangle_{1_c}|R(2,3)\rangle$,
which is BRST invariant, might not define a gauge invariant 
because it includes non-primary part 
$\partial^2 c c-\bar\partial^2\bar c \bar c$ in the closed string side.\\

In the following, we derive eigenvalues of the on-shell closed string
states for $K_n\equiv L_n-(-1)^nL_{-n}$
 mentioned in \S \ref{sec:On_Gauge},
through the open-closed string vertex $\langle \hat{\gamma}(1_{\rm
c},2)|$. Let us consider the CFT correlator on $z$-plane,
where each local unit disks are mapped by $z=h_r(w_r)$ ($r=1,1^*,2$).
We define a 1-form $\omega_n$ on $z$-plane by
\begin{eqnarray}
 \omega_n&\equiv&\frac{dz}{2\pi
  i}\frac{v_n(h^{-1}_2(z))}{(h_2^{-1})'(z)}T(z)\,,~~~~
v_n(w)=w(w^n-(-1)^nw^{-n})\,,
\end{eqnarray}
where $h_2^{-1}(z)$ is the inverse map of $h_2(w_2)=I\circ f_{\cal I}(w_2)$.
Using the explicit form of $h_2^{-1}(z)=z+\sqrt{z^2+1}$, $\omega_n$
can be rewritten as
\begin{eqnarray}
 \omega_n&=&\frac{dz}{2\pi i}(1+z^2)P_n(z)T(z)\,,\\
P_n(z)&=&\sum_{l=0}^{[(n-1)/2]}\frac{2\,n!}{(2l+1)!(n-2l-1)!}
z^{n-2l-1}(z^2+1)^l \,,
\end{eqnarray}
where $[(n-1)/2]$ is $(n-1)/2$ for odd $n$ and $n/2-1$ for even $n$.
Because $P_n(z)$ is a polynomial, $\omega_n$ is regular except for
$z=h_1(0)=i$, $z=h_{1^*}(0)=-i$ and $z=h_2(0)=\infty$, where 
some fields, which correspond to closed and open string 
states, are inserted in the correlator.
As in \cite{Rastelli:2000iu}, we deform
a contour, which gives trivial result $\oint \omega_n=0$
in the correlator, to encircle each origin of local coordinates $w_r$
($r=1,1^*,2$) and get a relation:
\begin{eqnarray}
 0&=&\langle \hat\gamma(1_{\rm
  c},2)|\sum_{r=1,1^*,2}\oint_{w_r=0}\omega_n\,.
\label{eq:0=oint}
\end{eqnarray}
Noting a transformation law of the energy momentum tensor:
\begin{eqnarray}
 T(z)&=&(h_r'(w_r))^{-2}\left(T^{(r)}(w_r)-\frac{c}{12}
S(h_r(w_r),w_r)\right),
\end{eqnarray}
where $c$ is the central charge of the associated Virasoro algebra
 and $S(z,w)$ is the Schwarzian derivative defined by
$S(f(z),z)=\frac{f'''(z)}{f'(z)}-\frac{3}{2}\left(
\frac{f''(z)}{f'(z)}\right)^2$, we can compute the right hand side of
 (\ref{eq:0=oint}). Using  $S(h_2(w_2),w_2)=6(1+w_2^2)^{-2}$, 
the residue around $w_2=0$, i.e., open string side, is computed as
\begin{eqnarray}
\langle \hat\gamma(1_{\rm c},2)|\oint_{w_2=0}\omega_{2m-1} &=&
\langle \hat\gamma(1_{\rm c},2)|K_{2m-1}^{(2)}\,,\\
\langle \hat\gamma(1_{\rm c},2)|\oint_{w_2=0}\omega_{2m} &=&
\langle \hat\gamma(1_{\rm c},2)|(K_{2m}^{(2)}-(-1)^mmc/2)\,.
\end{eqnarray}
In the closed string side, we can evaluate the residue as
\begin{eqnarray}
 \langle \hat\gamma(1_{\rm c},2)|\oint_{w_1=0}\omega_n &=&
\langle \hat\gamma(1_{\rm
c},2)|\left(4ni^nL_0^{(1)}+\sum_{k=2}^{\infty}f_{n,k}L_{k-1}^{(1)}\right)\,,\\
 \langle \hat\gamma(1_{\rm c},2)|\oint_{w_{1^*}=0}\omega_n &=&
\langle \hat\gamma(1_{\rm
c},2)|\left(4n(-i)^n\bar L_0^{(1)}
+\sum_{k=2}^{\infty}\bar f_{n,k}\bar L_{k-1}^{(1)}\right)\,,
\end{eqnarray}
where use has been made of $h_1(0)=i,h_1'(0)\neq 0$ and the coefficients
$f_{n,k}$ are determined by the expansion
\begin{eqnarray}
 \frac{1+h_1(w_1)^2}{h_1'(w_1)}P_n(h_1(w_1))&=&
4ni^nw_1+\sum_{k=2}^{\infty}f_{n,k}w_1^k\,.
\end{eqnarray}
Therefore, from (\ref{eq:0=oint}), we obtain the relations on the
vertex  $\langle \hat{\gamma}(1_{\rm c},2)|$:
\begin{eqnarray}
&&0= \langle \hat\gamma(1_{\rm c},2)|\biggl(\!
K_{2m-1}^{(2)}\!-4(2m\!-\!1)i(-1)^m(L_0^{(1)}\!-\bar L_0^{(1)} )
+\!\sum_{k=2}^{\infty}(f_{2m-1,k}L_{k-1}^{(1)}
\!+\!\bar f_{2m-1,k}\bar L_{k-1}^{(1)}\!
\biggr),\nonumber\\
\\
&&0= \langle \hat\gamma(1_{\rm c},2)|\biggl(\!
K_{2m}^{(2)}\!-(-1)^mm\frac{c}{2}+8m(-1)^m(L_0^{(1)}\!+\!\bar L_0^{(1)} )
+\!\sum_{k=2}^{\infty}(f_{2m,k}L_{k-1}^{(1)}
+\bar f_{2m,k}\bar L_{k-1}^{(1)}\!
\biggr).\nonumber\\
\end{eqnarray}
Using the above formulae and noting that $K_n$ is BPZ odd,
we can derive the eigenvalue of the closed string state in the open
string Hilbert space mapped from a $(h,\bar h)$-primary state $|h,\bar
h\rangle$ in the closed string Hilbert space:
\begin{eqnarray}
&&K_{2m-1}^{(3)}(\langle \hat\gamma(1_{\rm c},2)|h,\bar h\rangle_{1_{\rm
  c}}|R(2,3)\rangle)=-(\langle \hat\gamma(1_{\rm c},2)|K_{2m-1
}^{(2)}|h,\bar h\rangle_{1_{\rm
  c}})|R(2,3)\rangle\nonumber\\
&&=-4(2m-1)i(-1)^m(h-\bar h)
\langle \hat\gamma(1_{\rm c},2)|h,\bar h\rangle_{1_{\rm
  c}}|R(2,3)\rangle\,,
 \label{eq:K_nsym_gen_odd}
\\
&&K_{2m}^{(3)}(\langle \hat\gamma(1_{\rm c},2)|h,\bar h\rangle_{1_{\rm
  c}}|R(2,3)\rangle)=-(\langle \hat\gamma(1_{\rm c},2)|K_{2m
}^{(2)}|h,\bar h\rangle_{1_{\rm
  c}})|R(2,3)\rangle\nonumber\\
&&=(-1)^mm(8(h+\bar h)-c/2)
\langle \hat\gamma(1_{\rm c},2)|h,\bar h\rangle_{1_{\rm
  c}}|R(2,3)\rangle\,.
 \label{eq:K_nsym_gen_even}
\end{eqnarray}
In particular, for $c=0,h=\bar h=0$, the above relations
 yield (\ref{eq:symK_n}).
By taking the energy momentum tensor in the matter sector, 
we have (\ref{eq:symK_nmat})
with $c=26$ and $h=\bar h=1$.
For the energy momentum tensor in the ghost sector, 
we have (\ref{eq:symK_ngh}) 
with $c=-26$ and $h=\bar h=-1$.\\

In this section, we have used the Shapiro-Thorn's vertex $\langle
\hat\gamma(1_{\rm c},2)|$ for an explicit example.
However, it is not unique open-closed string vertex in order to get 
the gauge invariant overlap ${\cal O}_V(\Psi)$ (\ref{eq:O_Vgamma})
in open string field theory because we assume that
$V_{\rm c}(z,\bar z)$ is a primary field with dimension $(0,0)$.
Actually, we can take other map in the closed string side 
such as $h_1(0)=i,~h_1'(0)\ne 0$ in the definition of 
the vertex (\ref{eq:S-T_LPPdef})
to construct ${\cal O}_V(\Psi)$.

\section{$L_0$-Level Truncation Expansion of $\psi_r$
\label{sec:leveltr}
}

Let us investigate the  $L_0$-level truncation of $\psi_r$ (\ref{eq:psi_r})
because Schnabl's solution $\Psi_{\lambda}$ in \cite{Schnabl_tach}
is made of $\psi_r$ and its derivative with respect to $r$.
By moving ${\cal L}_0,{\cal B}_0$ to the right in  $\psi_{r-2}$, one gets
\begin{eqnarray}
&&\psi_{r-2}=\frac{1}{\pi}U^{\dagger}_r\left(
\frac{r}{\pi}{\cal B}^{\dagger}_0
{\tilde c}(-\tilde z_r)
{\tilde c}(\tilde z_r)
+{\tilde c}(-\tilde z_r)+{\tilde c}(\tilde z_r)\right)|0\rangle\,,
~~~~\tilde z_r\equiv \frac{\pi(r-2)}{2r}.
\end{eqnarray}
The state $|\chi_{r-2}\rangle$ is given by removing the Virasoro
operator $U^{\dagger}_r$,
 as $|\chi_{r-2}\rangle\equiv (U_r^{\dagger})^{-1}|\psi_{r-2}\rangle$.
Then, it should be expanded in terms of the ordinary
oscillators for the $bc$-ghosts as 
\begin{eqnarray}
{}|\chi_{r-2}\rangle
&=&\sum_{p\ge -1}g_pc_{-p}|0\rangle+\frac{1}{2}\sum_{s\ge 2}\sum_{p,q\ge
 -1}g_{s;p,q}b_{-s}c_{-p}c_{-q}|0\rangle\,.
\label{eq:chi_r-2}
\end{eqnarray}
Noting the relations
\begin{eqnarray}
&&\langle 0|c_{-1}c_0c_1b_pc_{-p'}|0\rangle=\delta_{p,p'}\,,~~~~~~
\langle 0|c_{-1}c_0c_1b_pb_{-s'}c_{-p'}c_{-q'}|0\rangle=0\,,\\
&&\langle
 0|c_{-1}c_0c_1b_qb_pc_sb_{-s'}c_{-p'}c_{-q'}|0\rangle
=2\delta_{s'}^s\delta_{p'}^{[p}\delta^{q]}_{q'}\,,~~~~~~
\langle
 0|c_{-1}c_0c_1b_qb_pc_sc_{-p'}|0\rangle=0,\\
&&\{b_p,\tilde c(\tilde z)\}=(\sin2\tilde z)(\tan\tilde z)^p/2\,,
~~~~{\cal
 B}^{\dagger}_0=b_0+\sum_{k=1}^{\infty}\frac{2(-1)^{k+1}}{4k^2-1}b_{-2k}\,,
\end{eqnarray}
($p,q,p',q'\ge -1$, $s,s'\ge 2$) in the ghost sector,
one can compute the coefficients in the expansion of $|\chi_{r-2}\rangle$:
\begin{eqnarray}
 g_p&=&\langle 0|c_{-1}c_0c_1b_p|\chi_{r-2}\rangle=\frac{1}{2\pi}(1-(-1)^p)
(\sin 2\tilde z_r)(\tan\tilde z_r)^p\left[1-\frac{r}{2\pi}\sin2\tilde
				     z_r\right],
\label{eq:gp_coeff}\\
g_{s;p,q}&=&\langle
 0|c_{-1}c_0c_1b_qb_pc_s|\chi_{r-2}\rangle\nonumber\\
&=&\frac{(1+(-1)^s)r}{4\pi^2}\frac{(-1)^{\frac{s}{2}+1}}{s^2-1}(\sin 2\tilde
 z_r)^2((-1)^q-(-1)^p)(\tan\tilde z_r)^{p+q}\,.
\label{eq:gpq_coeff}
\end{eqnarray}
The operator $U_r^{\dagger}
=(2/r)^{{\cal L}_0^{\dagger}}$
with ${\cal
L}_0^{\dagger}=L_0+\sum_{k=1}^{\infty}\frac{2(-1)^{k+1}}{4k^2-1}L_{-2k}$,
can be rewritten as
\begin{eqnarray}
U_r^{\dagger}&=&\cdots e^{u_{4}(r)L_{-4}}e^{u_{2}(r)L_{-2}}
\left(\frac{2}{r}\right)^{L_0}
\equiv
 \left[\prod_{k=1,\leftarrow}^{\infty}
e^{u_{2k}(r)L_{-2k}}\right]\left(\frac{2}{r}\right)^{L_0}\,.
\end{eqnarray}
Noting that $U_r$ is defined by the conformal map 
$F_r(z)=\tan\left((2/r)\arctan z\right)$ for the wedge state,
the coefficients $u_{2k}(r)$ ($k=1,2,\cdots$) are determined to be 
\begin{eqnarray}
u_2(r)&=&{1\over 3!}\left.\frac{d^3}{dz^3}\tilde F_r(z)\right|_{z=0}
=-\frac{(r+2)(r-2)}{3r^2}\,,
\label{eq:u2_formula}
\\
u_{2k}(r)&=&
{1\over (2k+1)!}\left.\frac{d^{2k+1}}{dz^{2k+1}}
f_{2k-2,-u_{2k-2}(r)}\!\circ\cdots
\circ\!f_{4,-u_4(r)}\!\circ\!f_{2,-u_2(r)}\!\circ\!\tilde
F_r(z)\right|_{z=0},\nonumber\\
&&(k\ge 2)
\label{eq:u2k_formula}
\\
\tilde F_r(z)&=&\frac{r}{2}F_r(z)=\frac{r}{2}
\tan\left(\frac{2}{r}\arctan z\right),\\
f_{n,t}(z)&=&
e^{tz^{n+1}\partial_z}z=\frac{z}{(1-tnz^n)^{1/n}}\,.
\end{eqnarray}
Using the above formulae, 
$\psi_{r-2}$ can be expressed in terms of the $L_0$-level as 
\begin{eqnarray}
 \psi_{r-2}&=&\left[\prod_{k=1,\leftarrow}^{\infty}
e^{u_{2k}(r)L_{-2k}}\right]\biggl[
\frac{1}{\pi}\sin\frac{2\pi}{r}\left(1-\frac{r}{2\pi}\sin\frac{2\pi}{r}
\right)\sum_{p\ge -1;p:{\rm odd}}\left(\frac{2}{r}\cot
\frac{\pi}{r}\right)^pc_{-p}|0\rangle
\label{eq:psi_r-2_level}\\
&&+\frac{r}{2\pi^2}\!\left(\sin\frac{2\pi}{r}\right)^2\!\!
\sum_{s\ge 2;s:{\rm even}}\!\!\frac{(-1)^{\frac{s}{2}+1}}{s^2-1}\!
\left(\!\frac{2}{r}\!\right)^s\!\!
\sum_{p,q\ge -1;p+q:{\rm odd}}\!\!(-1)^q\!\left(
\frac{2}{r}\cot\frac{\pi}{r}
\right)^{p+q}\!b_{-s}c_{-p}c_{-q}|0\rangle
\biggr].\nonumber
\end{eqnarray}
Note that $L_{-2n}$ in this formula is the total Virasoro
operator, which includes both of the matter and the ghost sector. 
For large $N$, $\psi_N$ behaves as
\begin{eqnarray}
 \psi_N&=&\frac{1}{N^3}\,\frac{4\pi^2}{3}
\left[\prod_{k=1,\leftarrow}^{\infty}\!
e^{u_{2k}(\infty)L_{-2k}}\right]\sum_{p\ge -1;p:{\rm
odd}}\left(\frac{2}{\pi}\right)^pc_{-p}|0\rangle
\label{eq:psi_largeN}
\\
&&+\frac{1}{N^3}\frac{8}{3}
\left[\prod_{k=1,\leftarrow}^{\infty}\!
e^{u_{2k}(\infty)L_{-2k}}\right]
\sum_{p,q\ge -1;p+q:{\rm odd}}\!(-1)^q\left(
\frac{2}{\pi}
\right)^{p+q}b_{-2}c_{-p}c_{-q}|0\rangle+{\cal O}(N^{-4}).\nonumber
\end{eqnarray}
Here, the coefficients $u_{2k}(\infty)$ ($k=1,2,\cdots$), which defines
the sliver state, are finite.
Therefore, $\psi_N$ behaves as $N^{-3}\to 0$ ($N\to \infty$) 
as long as one considers the inner product between $\psi_N$ and 
any ordinary Fock state. In this sense, the phantom term $\psi_{N+1}$
in the expression (\ref{eq:Psi_lambda_1_N}) of $\Psi_{\lambda=1}$ can be
ignored in the $L_0$-level truncation.

\section{Equivalence of Gauge Invariant Overlaps for
a Classical Solution in the Universal Fock Space
\label{sec:tachyon=dilaton}}

In this appendix, we will prove the equivalence of the gauge invariant overlaps 
for the classical solution in the universal Fock space. In particular, 
we will consider the gauge invariant overlaps of the closed string tachyon and
dilaton states. However, we can easily generalize this proof to
any other closed string state. 

For this purpose, we will consider an open string field 
of the form $\Psi_{\rm univ}=\sum_{L=0}^{\infty}\psi_L$ with
\begin{eqnarray}
\psi_L\!&=&\!\!\!\!
\sum_{p,q\ge 0,n_1\ge \cdots\ge  n_p\ge 2,j_q > \cdots > j_1\ge 1,
k_q > \cdots > k_1 \ge 0\atop 
n_1+\cdots +n_p+j_1+\cdots +j_l+k_1+\cdots +k_q=L
}\!\!\!\!\!\!
{\cal C}^{(L)}_{n_i,j_i,k_i}
L_{-n_1}^{({\rm m})}\cdots L_{-n_p}^{({\rm m})}
b_{-j_1}\cdots b_{-j_q}c_{-k_1}\cdots c_{-k_q}c_1|0\rangle,
\nonumber\\
\end{eqnarray}
where $L_{-n}^{({\rm m})}$ is the matter Virasoro operator
and the coefficients ${\cal C}^{(L)}_{n_i,j_i,k_i}$ are arbitrary
constants.
Namely, the matter part of $\Psi_{\rm univ}$ is expressed by the Virasoro
operator only and $\psi_L$ is the $L_0$-level $L$ sector of  $\Psi_{\rm
univ}$. Schnabl's solution $\Psi_{\lambda}$ has the same structure
as one can see from (\ref{eq:psi_r-2_level}) and (\ref{eq:sol2}).
For the tachyon state $\Phi_k$ in (\ref{eq:closed_tach})
and the massless state $\Phi_{\zeta}$ with zero momentum in 
(\ref{eq:closed_massless}), 
we will prove the equality  
\begin{eqnarray}
 \frac{1}{26}
\zeta_{MN}\eta^{MN}\langle \psi_L,\Phi_k\rangle&=&
\langle \psi_L,\Phi_{\zeta}\rangle
\label{eq:psi_L,k_zeta},~~~~(L=0,1,2,\cdots )
\end{eqnarray}
which implies that 
\begin{eqnarray}
 \langle\Phi_k,\Psi_{\lambda,L}\rangle
&=&\langle\Phi_{\eta},\Psi_{\lambda,L}\rangle
\label{eq:tach=dil}
\end{eqnarray}
for the case $\Psi_{\rm univ}=\Psi_{\lambda}$
and $\zeta_{MN}=\eta_{MN}$.

We denote the matter part of
$\Phi_k$ and $\Phi_{\zeta}$ as $\Phi_k^{({\rm m})}$ and 
 $\Phi_{\zeta}^{({\rm m})}$ respectively.
Because the ghost part of them is the same, which is given by
 $e^{E_{\rm gh}}c_0c_1|0\rangle_{\rm gh}$, (\ref{eq:psi_L,k_zeta}) is
 equivalent to
\begin{eqnarray}
 \frac{1}{26}
\zeta_{MN}\eta^{MN}
{}_{\rm mat}\langle 0| L_{n_p}^{({\rm m})}\cdots 
L_{n_1}^{({\rm m})}|\Phi_k^{({\rm m})}\rangle
={}_{\rm mat}\langle 0| L_{n_p}^{({\rm m})}\cdots 
L_{n_1}^{({\rm m})}|\Phi_{\zeta}^{({\rm m})}\rangle\,.
\label{eq:0LLL}
\end{eqnarray}
for any $p,n_1\ge \cdots\ge  n_p\ge 2$.
Using the explicit form of $\Phi_k$ in (\ref{eq:closed_tach}) and 
$\Phi_{\zeta}$ in (\ref{eq:closed_massless}),
the commutation relation $[L_n^{({\rm m})},\alpha_{-k}^M]=k\alpha_{n-k}^M$
and the identities given by \cite{Gross:1986ia} 
\begin{eqnarray}
&&K_{2n}^{({\rm m})}e^{E_{{\cal I},{\rm m}}}|0\rangle_{\rm mat}=
-13(-1)^n n\, e^{E_{{\cal I},{\rm m}}}|0\rangle_{\rm mat},~~~~~
 K_{2n-1}^{({\rm m})}e^{E_{{\cal I},{\rm m}}}|0\rangle_{\rm mat}=0\,,\\
&&K_n^{({\rm m})}\equiv L_n^{({\rm m})}-(-1)^nL_{-n}^{({\rm m})}\,,
\end{eqnarray}
($n=1,2,\cdots$)
for the matter sector of the identity state $|{\cal I}\rangle$
(\ref{eq:identity_osc}), one obtains
\begin{eqnarray}
 &&K_{2n}^{({\rm m})}|\Phi_k^{({\rm m})}\rangle
=(4\alpha'k^2-13)(-1)^n n|\Phi_k^{({\rm m})}\rangle,
~~~~K_{2n-1}^{({\rm m})}|\Phi_k^{({\rm m})}\rangle=0\,,
\label{eq:K_nPhi_k}\\
 &&K_{2n}^{({\rm m})}|\Phi_{\zeta}^{({\rm m})}\rangle
=3(-1)^n n|\Phi_{\zeta}^{({\rm m})}\rangle,
~~~~~~~~~~~~~~~~~~
K_{2n-1}^{({\rm m})}|\Phi_{\zeta}^{({\rm m})}\rangle=0\,.
\label{eq:K_nPhi_zeta}
\end{eqnarray}
Namely, both $\Phi_k^{({\rm m})}$ and  $\Phi_{\zeta}^{({\rm m})}$ 
are the eigenstates of the operators $K_n^{({\rm m})}$ and 
have the same eigenvalue with the on-shell condition $\alpha'k^2=4$
for the tachyon state. 
These relations are also expected from the argument in
(\ref{eq:K_nsym_gen_odd}) and (\ref{eq:K_nsym_gen_even}).
Hence, one finds that 
\begin{eqnarray}
 \frac{1}{26}
\zeta_{MN}\eta^{MN}
 {}_{\rm mat}\langle 0| K_{m_r}^{({\rm m})}\cdots 
K_{m_1}^{({\rm m})}|\Phi_k^{({\rm m})}\rangle
= {}_{\rm mat}\langle 0| K_{m_r}^{({\rm m})}\cdots 
K_{m_1}^{({\rm m})}|\Phi_{\zeta}^{({\rm m})}\rangle,
\label{eq:0KKK}
\end{eqnarray}
because the normalization is fixed by
${}_{\rm mat}\langle 0|\Phi_k^{({\rm m})}\rangle=\frac{1}{4}$ and 
${}_{\rm mat}\langle 0|\Phi_{\zeta}^{({\rm m})}\rangle=
 \frac{1}{26\cdot 4}\zeta_{MN}\eta^{MN}$.
One can see that
${}_{\rm mat}\langle 0| L_{n_p}^{({\rm m})}\cdots 
L_{n_1}^{({\rm m})}
$ in (\ref{eq:0LLL}) can be rewritten as a linear combination of 
${}_{\rm mat}\langle 0| K_{m_{p'}}^{({\rm m})}\cdots 
K_{m_1}^{({\rm m})}$ ($p'\le p$)
by using ${}_{\rm mat}\langle 0|L_{-n}^{({\rm m})}=0$
($n=-1,0,1,2,\cdots$) and the Virasoro algebra
$[L_n^{({\rm m})},L_m^{({\rm m})}]=(n-m)L_{n+m}^{({\rm
m})}+\frac{13}{6}n(n^2-1)\delta_{n,-m}$.
In fact, it can be proved by using the relation 
 $\langle 0|L_n^{({\rm m})}=\langle 0|K_n^{(\rm m)}$
($n> 1$) and
\begin{eqnarray}
&&\langle 0|K_{m_r}^{({\rm m})}\cdots 
K_{m_1}^{({\rm m})}L_n^{({\rm m})}\nonumber\\
&=&\langle 0|\biggl(K_{m_r}^{({\rm m})}\cdots 
K_{m_1}^{({\rm m})}K_n^{({\rm m})}
+\sum_{l=1}^r[K_{m_l}^{({\rm m})},L_{-n}^{({\rm m})}]
K_{m_r}^{({\rm m})}\cdots 
K_{m_{l+1}}^{({\rm m})}K_{m_{l-1}}^{({\rm m})}\cdots
K_{m_1}^{({\rm m})}\nonumber\\
&&+\sum_{r\ge k>l\ge 1}[K_{m_k}^{({\rm m})},[K_{m_l}^{({\rm
 m})},L_{-n}^{({\rm m})}]]
K_{m_r}^{({\rm m})}\cdots 
K_{m_{k+1}}^{({\rm m})}K_{m_{k-1}}^{({\rm m})}\cdots
K_{m_{l+1}}^{({\rm m})}K_{m_{l-1}}^{({\rm m})}\cdots
K_{m_1}^{({\rm m})}
\nonumber\\
&&+\cdots 
+\sum_{l=1}^r[K_{m_r}^{({\rm m})},[K_{m_{r-1}}^{({\rm m})},
\cdots [K_{m_{l+1}}^{({\rm m})},[K_{m_{l-1}}^{({\rm m})},\cdots [K_{m_1}^{({\rm
 m})},L_{-n}^{({\rm m})}]\cdots]]\cdots]]K_{m_l}^{({\rm m})}
\nonumber\\
&&+[K_{m_r}^{({\rm m})},[K_{m_{r-1}}^{({\rm m})},\cdots [K_{m_1}^{({\rm
 m})},L_{-n}^{({\rm m})}]\cdots]]\biggr)
\end{eqnarray}
for $(m_i,n> 1)$ recursively.
Therefore, (\ref{eq:0LLL}) is derived from (\ref{eq:0KKK}).

\end{document}